\documentclass[prb,twocolumn,showpacs,amsmath,amssymb,superscriptaddress]{revtex4-1}

\usepackage{graphicx}
\usepackage{dcolumn}
\usepackage{soul}
\usepackage{bm}
\usepackage{bbm}
\usepackage{ulem}
\usepackage{color}
\usepackage{colortbl}
\usepackage{epstopdf}
\begin{document}

\title{Unconventional superconductivity and Surface pairing symmetry in Half-Heusler Compounds}

\author{Qing-Ze Wang}
\affiliation{Department of Physics, The Pennsylvania State University, University Park, Pennsylvania 16802-6300, USA}
\author{Jiabin Yu}
\affiliation{Department of Physics, The Pennsylvania State University, University Park, Pennsylvania 16802-6300, USA}
\author{Chao-Xing Liu}\email{cxl56@psu.edu}
\affiliation{Department of Physics, The Pennsylvania State University, University Park, Pennsylvania 16802-6300, USA}

\date{\today}
\begin{abstract}
	Signatures of nodal line/point superconductivity \cite{kim2016,brydon2016}
have been observed in half-Heusler compounds, such as LnPtBi (Ln = Y, Lu). Topologically non-trivial band structures, as well as topological surface states, has also been confirmed by angular-resolved photoemission spectroscopy in these compounds\cite{liuz2016}. In this work, we present a systematical classification of possible gap functions of bulk states and surface states in half-Heusler compounds and the corresponding topological properties based on the representations of crystalline symmetry group. Different from all the previous studies based on four band Luttinger model, our study starts with the six-band Kane model, which involves both four p-orbital type of $\Gamma_8$ bands and two s-orbital type of $\Gamma_6$ bands. Although the $\Gamma_6$ bands are away from the Fermi energy,
our results reveal the importance of topological surface states, which originate from the band inversion between $\Gamma_6$ and $\Gamma_8$ bands, in determining surface properties of these compounds in the superconducting regime by combining topological bulk state picture and non-trivial surface state picture.
\end{abstract}
\maketitle

\section{Introduction}
Ternary half-Heusler compounds with XYZ compositions, where X, Y atoms are from transition or rare-earth metals and Z is from main-group element, have attracted a great deal of researchers' attention for their tunability of electronic band structures (band gap, spin-orbit coupling strength, etc) and multiple functionality\cite{yan2014}.
Topological insulator phases have been theoretically predicted in more than 50 compounds of half-Heusler family \cite{chadov2010,lin2010,xiao2010a,yu2017}, thus providing us a fertile ground to search for other topological phases.
Recent experiment of angle-resolved photoemission spectroscopy (ARPES) has observed topological surface states(TSSs) in half-Heusler compounds YPtBi and LuPtBi\cite{liuz2016}.
More interestingly, superconductivity has also been found to coexist with topologically non-trivial band structures
in these two half-Heusler compounds\cite{butch2011,PhysRevB.86.064515,tafti2013}. Experimentally,
the upper critical field $H_{c2}$ of YPtBi was found significantly higher than
the orbital limit\cite{butch2011} and the temperature-dependent penetration length in YPtBi
follows a power law behavior, instead of exponential dependence in normal s-wave superconductors\cite{kim2016}.
In addition, the first principles calculation has also excluded the conventional electron-phonon coupling mechanism for superconductivity in these compounds\cite{meinert2016}.
All these theoretical experimental results suggest the possibility of
unconventional superconductivity in these superconducting half-Heusler compounds.

In this work, we aim in a systematic classification of bulk and surface superconducting gap functions
in half-Heusler superconductors based on the six-band Kane model.
The previous theoretical studies based on
four band Luttinger model have suggested the possibility of quintet ($J=2$) and septet ($J=3$)
pairing functions due to spin-3/2 fermions \cite{brydon2016,timm2017,wu2006hidden,PhysRevLett.117.075301,boettcher2017unconventional},
and various topological superconducting phases with different pairing symmetries
\cite{kim2016, brydon2016, savary2017, roy2017, venderbos2017,PhysRevB.96.144514,yu2017singlet,PhysRevB.95.144503}.
However, the Luttinger model only takes into account four $\Gamma_8$ bands originating from the
p atomic orbitals, but neglects $\Gamma_6$ bands from the s atomic orbitals.
Although the $\Gamma_6$ bands are far away from the Fermi energy for YPtBi and LuPtBi,
the band inversion between the $\Gamma_6$ and $\Gamma_8$
bands can lead to TSSs, which can coexist with bulk superconductivity.
A recent experimental report indicates a significant higher transition temperature
at the surface than that in the bulk for LuPtBi \cite{banerjee2015}.
Thus, this leads to the interesting question how surface
superconductivity is related to bulk superconductivity and how it affects surface properties
in these half-Heusler superconductors.
This is the main question that we hope to understand in this work.

Below we will first present a systematic classification of possible bulk gap functions
for half-Heusler compounds based on the six-band Kane model in Section II. In Section III,
mirror symmetry protected topological superconductivity with non-trivial surfaces states
is demonstrated in several types of gap functions. Some results in this section have been known and the purpose of this section is to establish the bulk topological invariant which will reflect itself in the surface dispersion discussed in the next section.
In Section IV, we will project the bulk gap functions into the Hilbert space spanned by topological
surface states and establish the relationship between bulk gap functions and surface gap functions.
We will also study possible Majorana bands due to surface superconducting pairings.

\section{Model Hamiltonian and classification of gap function in $T_d$ group}
\begin{figure}[htb]
	\centering
	\includegraphics [width = 0.9\columnwidth,angle=0]{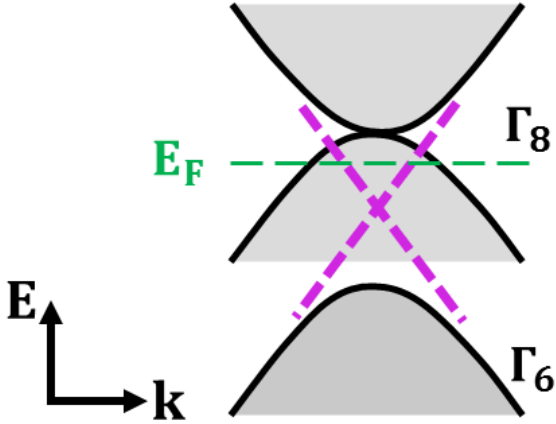}
	\caption{(Color online) Schematic diagram of band structure for half-Heusler compounds with inverted band structure. Four-fold degenerate $\Gamma_8$ bands is above two-fold degenerate $\Gamma_6$ band. The pink lines indicate the Dirac-cone TSSs for a finite-size system with an open boundary and the shaded region is the projected bulk states. The green dashed line is the Fermi level.}
    \label{fig1}
\end{figure}

Similar to the conventional zinc-blende semiconductors, the crystal structure of half-Heusler compounds possesses $T_d$
group symmetry and thus their low energy physics can be described by the eight-band Kane model\cite{novik2005,chadov2010,lin2010}.
The detailed form of the Kane Hamiltonian is shown in the Appendix A.
The inversion breaking term (C term in the $\Gamma_8$ part of Kane model) is not taken into account in our calculation.
Since $\Gamma_7$ band is far from the Fermi energy for LnPtBi (Ln = Lu, Y), we neglect $\Gamma_7$ band and only focus on two $\Gamma_6$ bands and four $\Gamma_8$ bands, which give rise to the six-band Kane model. $\Gamma_6$ band mainly consists of s-orbital and has a total angular momentum $j = \frac{1}{2}$ due to spin while $\Gamma_8$ bands, consisting of heavy-hole and light-hole bands, has a total angular momentum $j = \frac{3}{2}$ due to the combination of p-orbital angular momentum and spin. As shown in Fig. \ref{fig1}, the $\Gamma_6$ bands have a lower energy compared to the $\Gamma_8$ bands, thus forming an inverted band structure. This band inversion suggests the existence of TSSs coexisting and hybridizing with bulk states at the surface,
as schematically depicted in Fig. \ref{fig1}. Such surface states have recently been observed by ARPES experiment\cite{liuz2016}.


Next, we hope to implement a systematical classification of all possible gap functions for the six-band Kane model. $T_d$ point group can be generated by mirror symmetry operation with respect to (110) plane $M_{[110]}$, three-fold rotational operation along [111] direction $C_3$ and improper four-fold rotational operation along [001] direction $S_4$. Thus, we will use these three symmetry operations to classify possible s-wave on-site pairings for $\Gamma_6$ and $\Gamma_8$ band as well as the pairings between $\Gamma_6$ and $\Gamma_8$ bands (See Appendix B for more details). For Cooper pairs within $\Gamma_6$ bands, the possible s-wave pairing $\delta_s$ is the conventional spin-singlet pairing, given by $c_{\Gamma_{6},\uparrow}c_{\Gamma_{6},\downarrow} - c_{\Gamma_{6},\downarrow}c_{\Gamma_{6},\uparrow}$ or $\delta_s = i\xi_1\sigma_y$ in a matrix form, which belongs to $A_1$ ($\Gamma_1$) irreducible representation (IrRep) of $T_d$ group. Possible gap functions for the $\Gamma_8$ bands are listed in Table \ref{pairf}, from which one can see that the singlet gap function $\delta_1$ belong to $A_1$ IrRep, the quintet pairing functions $\delta_2$ and $\delta_4$ belong to the E IrRep and the quintet pairing functions $\delta_3$, $\delta_5$ and $\delta_6$ form a $T_2$ IrRep.
In addition, we also consider the p-wave septet pairing \cite{brydon2016} and d-wave quintet pairing\cite{yu2017singlet}, which are both in $A_1$ IrRep.
The p-wave septet pairing for the $\Gamma_8$ bands takes the form \cite{brydon2016}
\begin{eqnarray}
\delta_p  = \frac{\Delta_p}{4}\left(\begin{array}{cccc}
           3k_+&0&2\sqrt{3}k_z&-\sqrt{3}k_-\\
          0&-3k_-&\sqrt{3}k_+&2\sqrt{3}k_z\\
          2\sqrt{3}k_z&\sqrt{3}k_+&3k_-&0\\
          -\sqrt{3}k_-&2\sqrt{3}k_z&0&-3k_+
    \end{array}
	\right),
\end{eqnarray}
while the d-wave quintet pairing should be described by
\begin{equation}
\delta_d=\Delta_d \sum_{i=1}^5 g_{\mathbf{k},i}\Gamma^i
\end{equation}
with
$g_{\mathbf{k},1}=\sqrt{3} k_y k_z$,
$g_{\mathbf{k},2}=\sqrt{3} k_z k_x$,
$g_{\mathbf{k},3}=\sqrt{3} k_x k_y$,
$g_{\mathbf{k},4}=\frac{\sqrt{3}}{2} (k_x^2-k_y^2)$,
$g_{\mathbf{k},5}=\frac{1}{2}(2 k_z^2-k_x^2-k_y^2)$,
$\Gamma^1=\frac{1}{\sqrt{3}} (J_y J_z+J_z J_y)$,
$\Gamma^2=\frac{1}{\sqrt{3}} (J_z J_x+J_x J_z)$,
$\Gamma^3=\frac{1}{\sqrt{3}} (J_x J_y+J_y J_x)$,
$\Gamma^4=\frac{1}{\sqrt{3}} (J_x^2-J_y^2)$,
$\Gamma^5=\frac{1}{3} (2 J_z^2-J_x^2-J_y^2)$
and expressions of $J_{x,y,z}$ shown in Appendix \ref{cooper_pairings_bands_gamma8}.
The gap functions between the $\Gamma_6$ and $\Gamma_8$ bands fall into three IrReps, which are listed in Table \ref{pairf2}.

\begin{center}
\begin{table}[htb]
  \centering
  \begin{minipage}[t]{1.\linewidth}
  \centering
	  \caption{On-site Cooper pairings formed by states within $\Gamma_8$ bands and their classification. The third column list their matrix forms on the basis $\Psi_{\Gamma_8} = (|\Gamma_8, \frac{1}{2} \rangle,|\Gamma_8, -\frac{1}{2} \rangle,|\Gamma_8, \frac{3}{2} \rangle,|\Gamma_8, -\frac{3}{2} \rangle)^T$. Here $\tau_{x,y,z}$ and $\sigma_{x,y,z}$ are Pauli matrices. }
\hspace{-1cm}
\begin{tabular}
[c]{cccc}\hline\hline
                     & Cooper pair & $\Psi_{\Gamma_8}$&IrRep \\\hline
$\delta_1$ &$-c_{\Gamma_{81},\uparrow}c_{\Gamma_{81},\downarrow}+c_{\Gamma_{83},\uparrow}c_{\Gamma_{83},\downarrow} $ &$\tau_z\otimes i\sigma_y$& $A_1$[$\Gamma_1$]\\
\hline
$\delta_2$ &$-c_{\Gamma_{81},\uparrow}c_{\Gamma_{81},\downarrow} -c_{\Gamma_{83},\uparrow}c_{\Gamma_{83},\downarrow} $ &$\tau_0\otimes i\sigma_y$ & \\
$\delta_4$ &$-c_{\Gamma_{81},\uparrow}c_{\Gamma_{83},\uparrow} +c_{\Gamma_{81},\downarrow}c_{\Gamma_{83},\downarrow}$ &$i\tau_y\otimes \sigma_z$&E[$\Gamma_3$]\\\hline
$\delta_3$ &$-i(c_{\Gamma_{81},\uparrow}c_{\Gamma_{83},\uparrow} +c_{\Gamma_{81},\downarrow}c_{\Gamma_{83},\downarrow}) $ & $\tau_y\otimes \sigma_0$ &\\
$\delta_5$ &$-c_{\Gamma_{81},\uparrow}c_{\Gamma_{83},\downarrow} -c_{\Gamma_{81},\downarrow}c_{\Gamma_{83},\uparrow} $ &$i\tau_y\otimes \sigma_x$ &\\
$\delta_6$ &$i(-c_{\Gamma_{81},\uparrow}c_{\Gamma_{83},\downarrow} +c_{\Gamma_{81},\downarrow}c_{\Gamma_{83},\uparrow})$ & $\tau_x\otimes \sigma_y$&$T_2$[$\Gamma_4$] \\
\hline\hline
\label{pairf}
\end{tabular}
  \end{minipage}
\end{table}
\end{center}

The above discussion of pairing functions suggests that unconventional superconductivity could exist in half-Heusler compounds beyond spin singlet or triplet pairing. Since only $|\Gamma_8,\pm 3/2 \rangle$ states appear near the Fermi energy for a sample with hole doping, one needs to project the gap functions onto the bulk states $|\Gamma_8,\pm \frac{3}{2} \rangle$. To perform the projection, one may first diagonalize the Hamiltonian of the electronic states through an unitary matrix $U(k)$, i.e. $H_{eff}(\bold{k}) = U(\bold{k})^\dag H_e(\bold{k}) U(\bold{k})$, where $H_{eff}$ is a diagonal matrix with eigen-energies for different bands. The corresponding BdG Hamiltonian can be diagonlized by a unitary transformation $\tilde{U}(\bold{k}) = diag[U(\bold{k}), U(-\bold{k})^*]$. Since $
\tilde{U}(\bold{k})^\dag  H_{BdG} \tilde{U}(\bold{k}) =  \left (\begin{array}{cc}
		H_{eff}(\bold{k})- \mu&U(\bold{k})^\dag\Delta U(-\bold{k})^*\\
		U(-\bold{k})^T\Delta^\dag U(-\bold{k}) &- H_{eff}^*(-\bold{k}) + \mu
\end{array}\right)
$, the effective gap function can be obtained as
\begin{eqnarray}
\tilde{\Delta}(\bold{k}) = U(\bold{k})^\dag\Delta(\bold{k}) U(-\bold{k})^*.
\label{pairing_projection}
\end{eqnarray}
The projected gap functions of Cooper pairings onto $|\Gamma_8, \pm \frac{3}{2} \rangle$ can be decomposed as a function of spherical harmonics in the form
\begin{eqnarray}
\nonumber \tilde{\Delta}(\bold{k})  =\left\lbrace  \begin{array}{c}  \sum_m c_m Y_{lm}(\bold{k}) i\sigma_y, \qquad l = even \\
\sum_{m, \hat{n} = x,y,z} c_{\hat{n},m} Y_{lm}(\bold{k}) \hat{\sigma}\cdot \hat{n} i\sigma_y, \qquad l = odd \end{array}\right.,
\end{eqnarray}
where $c_m, c_{\hat{n},m}$ are complex coefficients and $Y_{lm}(\bold{k})$ are the spherical harmonics. The major components of the projected gap functions are listed in Table \ref{gapproj}, from which one can see that the projected $\delta_{s,1,d}$ behaves as a s-wave gap function, the projected $\delta_{p,7,8,12}$ behave as p-wave gap functions and $\delta_{2,3,4,5,6}$ behave as d-wave gap function.
After the projection, the coefficients $\delta_{9,10,11,13,14}$ are negligible on the Fermi surface and we do not discuss them here.

\begin{widetext}
\begin{center}
\begin{table}[htb]
  \centering
  \begin{minipage}[t]{1.\linewidth}
  	\centering
	  \caption{On-site Cooper pairings formed from states between $\Gamma_6$ and $\Gamma_8$ bands and their classification.}
\hspace{-1cm}
\begin{tabular}
[c]{cccc}\hline\hline
                     & Cooper pair & &IrRep \\\hline
$\delta_{11}$ &$-c_{\Gamma_{6},\uparrow}c_{\Gamma_{81},\downarrow} - c_{\Gamma_{6},\downarrow}c_{\Gamma_{81},\uparrow} $ &$\left(
	\begin{array}{cccc}
          0&1&0&0\\
          1&0&0&0
    \end{array}
	\right)$&\\
$\delta_{13}$ &$-c_{\Gamma_{6},\uparrow}c_{\Gamma_{83},\uparrow} - c_{\Gamma_{6},\downarrow}c_{\Gamma_{83},\downarrow} $ &$\left(
	\begin{array}{cccc}
          0&0&1&0\\
          0&0&0&1
    \end{array}
	\right)$& E[$\Gamma_3$]\\ \hline
$\delta_7$ &$\frac{i}{2}(c_{\Gamma_{6},\uparrow}c_{\Gamma_{81},\uparrow} +c_{\Gamma_{6},\downarrow}c_{\Gamma_{81},\downarrow}) -  \frac{i\sqrt{3}}{2}(c_{\Gamma_{6},\uparrow}c_{\Gamma_{83},\downarrow} +c_{\Gamma_{6},\downarrow}c_{\Gamma_{83},\uparrow})$ &$\frac{i}{2}\left(
	\begin{array}{cccc}
          1&0&0&-\sqrt{3}\\
          0&1&-\sqrt{3}&0
    \end{array}
	\right)$ & \\
$\delta_8$ &$-\frac{1}{2}(c_{\Gamma_{6},\uparrow}c_{\Gamma_{81},\uparrow} -c_{\Gamma_{6},\downarrow}c_{\Gamma_{81},\downarrow}) -  \frac{\sqrt{3}}{2}(c_{\Gamma_{6},\uparrow}c_{\Gamma_{83},\downarrow} -c_{\Gamma_{6},\downarrow}c_{\Gamma_{83},\uparrow})$&$\frac{1}{2}\left(
	\begin{array}{cccc}
          1&0&0&\sqrt{3}\\
          0&-1&-\sqrt{3}&0
    \end{array}
	\right)$ &\\
$\delta_{12}$ & $i(-c_{\Gamma_{6},\uparrow}c_{\Gamma_{81},\downarrow} + c_{\Gamma_{6},\downarrow}c_{\Gamma_{81},\uparrow})$& $\left(
	\begin{array}{cccc}
          0&-i&0&0\\
          i&0&0&0
    \end{array}
	\right)$&$T_2$[$\Gamma_4$]\\\hline
$\delta_9$ &$-\frac{i\sqrt{3}}{2}(c_{\Gamma_{6},\uparrow}c_{\Gamma_{81},\uparrow} +c_{\Gamma_{6},\downarrow}c_{\Gamma_{81},\downarrow}) -  \frac{i}{2}(c_{\Gamma_{6},\uparrow}c_{\Gamma_{83},\downarrow} +c_{\Gamma_{6},\downarrow}c_{\Gamma_{83},\uparrow})$ & $\frac{i}{2}\left(
	\begin{array}{cccc}
          -\sqrt{3}&0&0&-1\\
          0&-\sqrt{3}&-1&0
    \end{array}
	\right)$& \\
$\delta_{10}$ &$-\frac{\sqrt{3}}{2}(c_{\Gamma_{6},\uparrow}c_{\Gamma_{81},\uparrow} -c_{\Gamma_{6},\downarrow}c_{\Gamma_{81},\downarrow}) -  \frac{1}{2}(-c_{\Gamma_{6},\uparrow}c_{\Gamma_{83},\downarrow} +c_{\Gamma_{6},\downarrow}c_{\Gamma_{83},\uparrow})$ &$\frac{1}{2}\left(
	\begin{array}{cccc}
          \sqrt{3}&0&0&-1\\
          0&-\sqrt{3}&1&0
    \end{array}
	\right)$&\\
$\delta_{14}$ & $i(c_{\Gamma_{6},\uparrow}c_{\Gamma_{83},\uparrow} - c_{\Gamma_{6},\downarrow}c_{\Gamma_{83},\downarrow})$&$\left(
	\begin{array}{cccc}
          0&0&i&0\\
          0&0&0&-i
    \end{array}
	\right)$& $T_1$[$\Gamma_5$]\\\hline
\hline
\label{pairf2}
\end{tabular}
  \end{minipage}
\end{table}
\end{center}

\begin{center}
\begin{table}[htb]
  \centering
  \begin{minipage}[t]{1.\linewidth}
  \centering
  \caption{Projected gap functions on the bulk state $\vert \Gamma_8,\pm \frac{3}{2} \rangle$ on the Fermi level to the leading angular momentum $l$. $e_{i}$ and $o_{i}$ are projected coefficients.
  }
\label{gapproj}
\hspace{-1cm}
\begin{tabular}
[c]{c|c|c}\hline\hline
$\delta_i$     &Irrep\\\hline
$\delta_s$ &$A_1$[$\Gamma_1$]&$e_s Y_{0,0}i\sigma_y$\\ \arrayrulecolor{green}\hline\arrayrulecolor{black}
$\delta_1$ &$A_1$[$\Gamma_1$]&$e_1 Y_{0,0}i\sigma_y$\\
\arrayrulecolor{black}\hline\hline
$\delta_2$ & &$e_2 Y_{2,0}i\sigma_y $\\
$\delta_4$ & E[$\Gamma_3$]&$e_4 (Y_{2,2}+Y_{2,-2})i\sigma_y$\\\arrayrulecolor{green}\hline\arrayrulecolor{black}
$\delta_{11}$ &&-\\
$\delta_{13}$ & E[$\Gamma_3$]&-\\ \arrayrulecolor{black}\hline\hline
$\delta_3$  &&$e_3 (-i Y_{2,2} + iY_{2,-2})i\sigma_y$\\
$\delta_5$& &$e_5 (Y_{2,1} -Y_{2,-1})i\sigma_y$\\
$\delta_6$  & $T_2$[$\Gamma_4$]&$e_6 (iY_{2,1} +iY_{2,-1})i\sigma_y$\\\arrayrulecolor{green}\hline\arrayrulecolor{black}

$\delta_7$ &&$o_{7} (iY_{1,1}+iY_{1,-1})\sigma_x + o'_{7} iY_{1,0}\sigma_0$\\
$\delta_8$&&$o_{8} (Y_{1,1}-Y_{1,-1})\sigma_x + o'_{8} Y_{1,0}\sigma_z$\\
$\delta_{12}$&$T_2$[$\Gamma_4$]&$o_{12} (iY_{1,1}(\sigma_0 + \sigma_z)/2 -iY_{1,-1}(\sigma_0-\sigma_z)/2)$\\\arrayrulecolor{black}\hline\hline

$\delta_9$ &&-\\
$\delta_{10}$ &&-\\
$\delta_{14}$&$T_1$[$\Gamma_5$] &-\\
\hline\hline
 $\delta_{d}$ &$A_1$[$\Gamma_1$]&$e_{d} Y_{0,0}i\sigma_y$\\
 \arrayrulecolor{green}\hline\arrayrulecolor{black}
$\delta_p$ & $A_1$[$\Gamma_1$]&$o_{p}[ Y_{1,1}(\sigma_0 + \sigma_z)/2 +Y_{1,-1}(\sigma_0-\sigma_z)/2]$\\
\hline\hline
\end{tabular}
  \end{minipage}
\end{table}
\end{center}
\end{widetext}

\section{Topological mirror superconductivity in half-Heusler compounds}
The above analysis has classified all possible pairing functions for half-Heusler superconductors
with on-site pairings in the six-band Kane model. The coexistence of superconductivity and nontrivial electronic band structure in half-Heusler compounds, LnPtBi (Ln = Y, Lu), implies the possibility of topological superconductivity\cite{chadov2010,liuz2016,tafti2013,butch2011}. This section will focus on topological mirror superconductor phase due to the existence of mirror symmetry in half-Heusler crystals. Some results in this section have been studied in the literature\cite{timm2017}, but to be self-contained, we will describe the topological invariant of the topological mirror superconductor phase, which will reflect itself in the surface state spectrum discussed in the next section.

\subsection{Mirror symmetry and BdG Hamiltonian}
For convenience, we first transform the Kane model into a new set of basis wavefunctions that are eigenmodes of mirror symmetry.
The mirror symmetry operator is defined as $\mathcal{M}_{[110]} \equiv C_2 \mathcal{I}$, where $C_2$ is the two-fold rotational symmetry operator along [110] direction and $\mathcal{I}$ is the inversion symmetry operator. For the $\Gamma_6$ bands, $C_{2,\Gamma_6} = e^{i\frac{(s_x+s_y)\pi}{\sqrt{2}}}$ on the basis function  $\Psi_{\Gamma_6} = (|\Gamma_6, \frac{1}{2} \rangle,|\Gamma_6, -\frac{1}{2} \rangle)$, where $s_{x,y} = \frac{1}{2}\sigma_{x,y}$. For the $\Gamma_8$ band, $C_{2,\Gamma_8} = e^{i\frac{(J_x+J_y)\pi}{\sqrt{2}}}$ on basis function $\Psi_{\Gamma_8} = (|\Gamma_8, \frac{1}{2} \rangle,|\Gamma_8, -\frac{1}{2} \rangle,|\Gamma_8, \frac{3}{2} \rangle,|\Gamma_8, -\frac{3}{2} \rangle)^T$.
$\mathcal{I} = 1(-1)$ is used as the inversion operator for $\Gamma_{6,(8)}$ band. Consequently, the mirror symmetry should take the form $\mathcal{M}_{[110]} = i\frac{\sqrt{2}}{2}diag[\sigma_x+\sigma_y,\sigma_x+\sigma_y,-\sigma_x+\sigma_y]$
on the basis functions $\Psi = (\Psi_{\Gamma_6},\Psi_{\Gamma_8})^T$.
One can easily check that $\mathcal{M}^2_{[110]} = -1$ and thus the corresponding mirror parities should be taken as $\pm i$. Furthermore, we rotation the momentum ($k_x$,$k_y$,$k_z$) to ($k_\perp$, $k_\parallel$, $k_z$), where $\hat{k}_{\perp} \equiv \frac{1}{\sqrt{2}}(\hat{k}_x + \hat{k}_y)$ is normal to the mirror invariant plane and $\hat{k}_{\parallel} \equiv \frac{1}{\sqrt{2}}(\hat{k}_y - \hat{k}_x)$ lies in the mirror invariant plane. Since the mirror symmetry operator $\mathcal{M}_{[110]}$ and the Hamiltonian commute with each other at the mirror invariant plane with $k_{\perp} = 0$, one can block-diagonalize the Hamiltonian into two separate subblock Hamiltonians, each of which has a definite mirror parity.

In the mirror parity $+i$ subspace, the new basis functions are written as $\Phi_{+i}= e^{i5\pi/8}(-\frac{1-i}{2}\vert \Gamma_6,1/2 \rangle - \frac{\sqrt{2}}{2}\vert \Gamma_6,-1/2 \rangle,\frac{1-i}{2}\vert \Gamma_8,1/2 \rangle + \frac{\sqrt{2}}{2}\vert \Gamma_8,-1/2 \rangle,\frac{\sqrt{2}}{2}\vert \Gamma_8,3/2 \rangle + \frac{-1+i}{2}\vert \Gamma_8,-3/2 \rangle)^T$. The Hamiltonian on this basis function reads
\begin{eqnarray}
&&e_{+i}(k_{\parallel},k_z)=\left(
	\begin{array}{ccc}
          M_6&-\frac{1}{\sqrt{6}}P(2k_z - ik_{\parallel})&-\frac{1}{\sqrt{2}} Pk_{\parallel}\\
          &M_{81}& -R + S_p \\
          h.c.& &M_{83}\\
    \end{array}
\label{eq:+i}
	\right)
\end{eqnarray}
where $M_6 = T = E_c + \frac{\hbar^2}{2m_0}(2F+1)(k^2_z + k^2_{\parallel})$, $M_{81} = U-V$ with $U = E_v - \frac{\hbar^2}{2m_0}\gamma_1(k^2_z + k^2_{\parallel})$ and $V = - \frac{\hbar^2}{2m_0}\gamma_2(-2k^2_z + k^2_{\parallel})$, $M_{83} = U+V$, $R = i\frac{\hbar^2}{2m_0} \sqrt{3}\gamma_3 k^2_{\parallel}$ and $S_P = -\frac{\hbar^2}{2m_0} 2\sqrt{3}\gamma_3k_{\parallel}k_z$. Here $R^* = -R$ is used.

Similarly, the Hamiltonian block with mirror parity $-i$ can be written as
\begin{eqnarray}
&&e_{-i}(k_{\parallel},k_z)=\left(
	\begin{array}{ccc}
          M_6&\frac{1}{\sqrt{6}}P(2k_z + ik_{\parallel})&\frac{1}{\sqrt{2}} Pk_{\parallel}\\
          &M_{81}& R + S_p \\
          h.c.& &M_{83}\\
    \end{array}
\label{eq:-i}
	\right),
\end{eqnarray}
on the basis function $\Phi_{-i}= e^{-i 5\pi/8}(\frac{\sqrt{2}}{2}\vert \Gamma_6,1/2 \rangle - \frac{1+i}{2}\vert \Gamma_6,$ $-1/2 \rangle,\frac{\sqrt{2}}{2}\vert \Gamma_8,1/2 \rangle - \frac{1+i}{2}\vert \Gamma_8,-1/2 \rangle,$ $\frac{1+i}{2}\vert \Gamma_8,3/2 \rangle + \frac{\sqrt{2}}{2}\vert \Gamma_8,-3/2 \rangle)^T$.

The above two block Hamiltonians $e_{\pm i}$ are related to each other by time reversal (TR) symmetry, whose TR operator is expressed as $\mathcal{T}= diag[ -i\sigma_yK, i\sigma_yK, -i\sigma_yK]$ on the basis $\Psi = (\Psi_{\Gamma_6},\Psi_{\Gamma_8})^T$ with the complex conjugate operator K. On the new basis functions $\Phi = (\Phi_{+i},\Phi_{-i})^T$, the TR operator is expressed as $\mathcal{T}' = -i\sigma_y \otimes \mathcal{I}_{3 \times 3}K$ and mirror symmetry operator is $\mathcal{M}'_{[110]} = i \sigma_z \otimes \mathcal{I}_{3 \times 3}$.

Next we focus on the resulting Bogoliubov-de Gennes (BdG) type of Hamiltonian on the mirror invariant plane, which is written as
\begin{eqnarray}
\nonumber H
= \frac{1}{2}\sum_{\bold{k}}(c_{+i}^{\dag}(\bold{k}), c_{-i}^{\dag}(\bold{k}), c_{+i}^T(-\bold{k}),c_{-i}^T(-\bold{k}))  H_{BdG} \left(\begin{array}{c} c_{+i}(\bold{k})\\c_{-i}(\bold{k}) \\c^{\dag T}_{+i}(-\bold{k})\\c^{\dag T}_{-i}(-\bold{k}) \end{array}\right )
\label{ham}
\end{eqnarray}
with
\begin{eqnarray}
H_{BdG} = \left (\begin{array}{cc}
		H_0(\bold{k})&\Delta(\bold{k})\\
		h.c. &-H_0^*(-\bold{k})
\end{array}\right),
\label{hbdg}
\end{eqnarray}
where $H_0(\bold{k}) =  \left( 	
\begin{array}{cc}
e_{+i}(\bold{k})- \mu&0\\
0&e_{-i}(\bold{k})- \mu
\end{array}
\right)$, $\bold{k} = (k_z, k_{||})$, $\mu$ is the chemical potential and $\Delta$ denotes the superconducting gap function
with the form
\begin{eqnarray}
&&\Delta(\bold{k}) = \left( 	
\begin{array}{cc}
\Delta^{(-)}_{+i}(\bold{k})&\Delta^{(+)}_{+i}(\bold{k})\\
\Delta^{(+)}_{-i}(\bold{k})&\Delta^{(-)}_{-i}(\bold{k})
\end{array}
\right).
\end{eqnarray}
Here $c_{\pm i}(\bold{k})$ is the annihilation operator on the basis $\Phi_{\pm i}$.
Each block in $H_0$ and $\Delta$ represents a 3 by 3 matrix.
The BdG Hamiltonian satisfies the particle-hole(PH) symmetry $C H_{BdG}(\bold{k}) C^{-1} = - H_{BdG}(-\bold{k})$ with the PH symmetry operator $C = \tau_x \otimes \mathcal{I}_{6 \times 6} K$, where $\tau_x$ acts on the Nambu space. Moreover, the PH symmetry (or Fermi statistics) requires the constraint $\Delta(\bold{k}) = - \Delta^T(-\bold{k})$ for the gap function.


Mirror symmetry permitted gap functions should behave as  $ \mathcal{M}'_{[110]}  \Delta  \mathcal{M}'^T_{[110]} = \eta \Delta $ with $\eta = \pm 1$. As a result, one can define a new mirror operator in the Nambu space for the BdG Hamiltonian in Eq. \ref{ham} on the mirror invariant plane $k_{\perp} = 0$ as
\begin{eqnarray}
\tilde{\mathcal{M}}_{\eta} = \left (\begin{array}{cc}
		\mathcal{M}'_{[110]} &\\
		&\eta \mathcal{M}'^*_{[110]}
\end{array}\right).
\label{mirror}
\end{eqnarray}
One can easily check that $\tilde{\mathcal{M}'}_{\eta} H_{BdG}(k_z,k_{\parallel}) \tilde{\mathcal{M}'}^{-1}_{\eta}  = H_{BdG}(k_z,k_{\parallel})$ for $\mathcal{M}'_{[110]} \Delta \mathcal{M}'^T_{[110]} = \eta \Delta$ on the mirror invariant momentum plane $k_{\perp} = 0$, where $\mathcal{M}'^\dag_{[110]} = \mathcal{M}^{'-1}_{[110]}$ is used. For $\eta=+ (-)$, the corresponding gap function, as well as the superconducting ground state wave function, is even (odd) under mirror symmetry operation. Thus, these two different mirror symmetry operators $\tilde{\mathcal{M}}_{\pm}$ classify two different types of gap functions.

\begin{figure}[tb]
\centering
    \includegraphics[width = 0.9\columnwidth,angle=0]{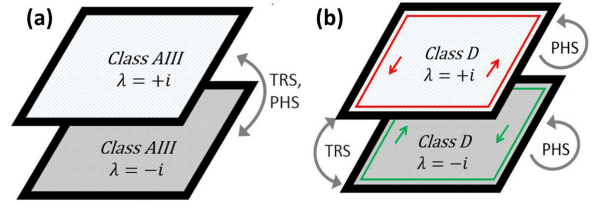}
    \caption{(Color online) (a), Configuration $\tilde{\mathcal{M}}_+$. PH symmetry interchanges the two mirror parity subspaces. The symmetry class for each subspace is AIII. (b), Configuration $\tilde{\mathcal{M}}_-$. PH symmetry exists in each mirror parity subspace. The symmetry class for each subspace is D. The red and green lines along the edge denote chiral edge modes in each mirror parity subspace.
    }
    \label{fig2}
\end{figure}

For the case of $\tilde{M}_+$ (even mirror parity pairing function), the gap function takes the form
\begin{eqnarray}
&&\Delta^{(+)} = \left( 	
\begin{array}{cc}
0 &\Delta^{(+)}_{+i}\\
\Delta^{(+)}_{-i}&0
\end{array}
\right),
\label{gap+}
\end{eqnarray}
on the basis set by the operators $\psi_{BdG} = (c_{+i}(k), c_{-i}(k),c^\dag_{+i}(-k), c^\dag_{-i}(-k))^T$,
where the superscript denotes the value of $\eta$ and the subscript $\pm i$ denotes mirror parity.
With the gap function Eq. (\ref{gap+}), the BdG Hamiltonian (\ref{hbdg}) can be rewritten into a block diagonal form with two blocks in the mirror parity $\pm i$ subspace
(Here the mirror parity $\pm i$ refers to the electron part set by the operators $c_{\pm i}(k)$).
While there is no TR symmetry and PH symmetry in each block Hamiltonian,
these two blocks are related by both TR and PH symmetry.
As a result, chiral symmetry $\Pi$, which is defined as $\Pi = C \times \mathcal{T}$\cite{schnyder2008},
exists in each block.
As a consequence, each block Hamiltonian with a fixed mirror parity belongs to the AIII symmetry class,
as illustrated in Fig. \ref{fig2}(a).
For the AIII symmetry class, there is no topological classification in 2D, but $\mathcal{Z}$ topological invariant in 1D
\cite{schnyder2008,schnyder2011,tewari2012}.
One can find a unitary matrix $V$ to transform the Hamiltonian into an off-block-diagonal form\cite{schnyder2011},
$VH_{BdG}V^\dag = \left( \begin{array} {cc} 0&q(\bold{k})\\q^\dag(\bold{k})&0\\ \end{array} \right)$ with $q^T(-\bold{k}) = -q(\bold{k})$ and the corresponding topological invariant (winding number) is defined as
\begin{eqnarray}
\nu = \frac{1}{2\pi i} \oint_{\textit{L}} dk_y Tr[q^{-1}(\bold{k})\nabla_{k_y}q(\bold{k})].
\label{winding2}
\end{eqnarray}

For the case of $\tilde{M}_-$ (odd parity pairing function), the gap function reads
\begin{eqnarray}
&&\Delta^{(-)} = \left( 	
\begin{array}{cc}
\Delta^{(-)}_{+i}&0\\
0&\Delta^{(-)}_{-i}
\end{array}
\right),
\end{eqnarray}
where the subscript and superscript are defined in the same way as those for $\Delta^{(+)}$.
In this case, the BdG Hamiltonian also takes a block diagonal form in
the mirror parity $\pm i$ subspace with each block preserving PH symmetry and two blocks related by TR symmetry.
Thus, each block Hamiltonian belongs to the D symmetry class and the corresponding topological invariant is
$\mathcal{Z}$ in 2D, defined by Chern number\cite{thouless1982,sinitsyn2006} and $\mathcal{Z}_2$ in 1D\cite{schnyder2008,schnyder2011,tewari2012}.

\subsection{Topological phases on mirror invariant planes}
The above symmetry analysis of gap functions suggests the possibility of TSC phase in both the gap functions
$\Delta^{(+)}$ and $\Delta^{(-)}$.
In this section, we will study TSC phases for the model Hamiltonian (4.5) explicitly for both cases.

\subsubsection{Even mirror parity case $\tilde{\Delta^{(+)}}$}
Since the Hamiltonian is block diagonal, we focus on the block part with mirror parity $+i$
while the block with mirror parity $-i$ can be related by TR symmetry.
The gap function $\Delta^{(+)}_{+i}$ for the $+i$ block takes the form (See Appendix C for more details)
\begin{eqnarray}
&&\Delta^{(+)}_{+i} = \left( 	
\begin{array}{ccc}
\xi_1&-i\xi_4+\xi_5&-\xi_8-i\xi_9\\
i\xi_4+\xi_5&-\xi_2&i\xi_6+\xi_7\\
-\xi_8+i\xi_9&-i\xi_6+\xi_7&\xi_3
\end{array}
\right).
\label{gap-m+}
\end{eqnarray}
on the basis set by the operators $\psi^{(+)} = (c_{+i}(\bold{k}), c^\dag_{-i}(-\bold{k}))^T$.
Here the gap functions are characterized by the parameters $\xi_i$ and their relationship
to $\delta_i$ used in the section II is listed in Table \ref{correspondence} in Appendix \ref{correspondence_delta_xi}.
The resulting Hamiltonian in the mirror parity $+i$ subspace is expressed as
\begin{eqnarray}
&&H^{(+)}_{+i} = \left( 	
\begin{array}{cc}
e_{+i}(\bold{k})&\Delta^{(+)}_{+i}\\
h.c.&-e^*_{-i}(-\bold{k})
\end{array}
\right)
\end{eqnarray}
where $e_{+i}$ and $e_{-i}$ are $3\times 3$ matrices, defined in Eq. \ref{eq:+i} and \ref{eq:-i}.
Since this Hamiltonian respects the chiral symmetry $\Pi^{(+)}_{+i} = -i \tau_y \otimes \mathcal{I}_{3 \times 3}$,
it can be transformed into a off-block diagonal form $H^{(+)'}_{+i} = \left ( 	
\begin{array}{cc}
0&q(\bold{k})\\
q^{\dag}(\bold{k})&0
\end{array}
\right )$ by a unitary transformation, where
\begin{widetext}
\begin{eqnarray}
q(\bold{k}) = \left(
\begin{array}{ccc}	
          M_6- \mu+i\xi_1&\frac{1}{\sqrt{6}}P(2k_z-ik_{\parallel}) + \xi_4 + i\xi_5&\frac{1}{\sqrt{2}} Pk_{\parallel} - i\xi_8 + \xi_9\\
          \frac{1}{\sqrt{6}}P(2k_z+ik_{\parallel}) - \xi_4 + i\xi_5&M_{81}-\mu - i\xi_2& R^* +S_P-\xi_6 + i\xi_7 \\
          \frac{1}{\sqrt{2}} Pk_{\parallel} - i\xi_8 - \xi_9& R +S_P^*+\xi_6 + i\xi_7&M_{83}-\mu+i\xi_3\\
\end{array}
\right).
\end{eqnarray}
\end{widetext}

\begin{figure}[tb]
    \includegraphics[width =1\columnwidth,angle=0]{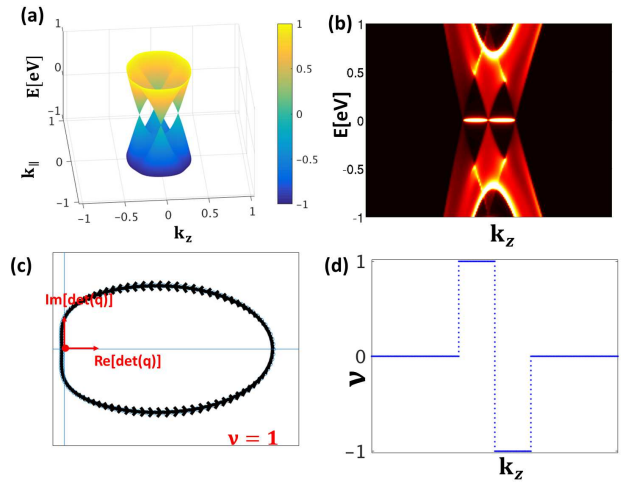}
    \caption{(Color online) (a) Bulk band structure as function of $k_z$ and $k_{||}$. (b) Surface DOS with $\hat{k}_{||}$ as the open boundary direction for $\tilde{M}_+$ configuration. The chemical potential is $\mu = -0.5$ eV. (c) Winding of $det(q(k_z = 0.1))$ around the origin on the complex plane. (d) Winding number as a function of $k_z$.}
    \label{fig3}
\end{figure}

TSC phase can exist for the above Hamiltonian, with the topological invariant defined by Eq. (\ref{winding2}).
As an example, we consider the gap function $\xi_7$, which belongs to $T_2$ IrRep.
The bulk energy dispersion of superconducting states is shown in Fig. \ref{fig3}(a) for the chemical potential
lying between the $\Gamma_8$ and $\Gamma_6$ bands, as shown in Fig. \ref{fig1}, and other parameters shown
in Table \ref{para_kane1}. One can see four nodes are present, thus giving rise to nodal superconductivity.
To further explore topological property of this nodal superconductivity,
we calculate the local density of states (LDOS) at the surface based on applying the
iterative Green's function method\cite{sancho1984} to the BdG Hamitlonian $H^{(+)}_{+i}$
in a semi-infinite system with an open boundary
along the $\hat{k}_{||}$ direction. 
As shown in Fig. \ref{fig3} (b), a zero-energy flat band appears between two bulk nodal points along $k_z$ direction.
To extract the topological nature of this zero-energy flat band,
we treat $k_z$ as a parameter and the bulk bands can be viewed as a 1D system with the momentum $k_{||}$.
As a result, the zero-energy band is protected by the winding number defined in Eq. (\ref{winding2}).
The winding number at a specific momentum $k_z$ can be simplified as $\nu_{k_z} = \frac{1}{2\pi i} \oint_{\textit{L}} d[ln(det (q(k_z,k_{||})))]$, where the integral loop is along $k_{||}$ from $-\pi$ to $\pi$. Fig. \ref{fig3} (c) shows the winding of $det (q(k_z,k_{||})))$ around the origin in the complex plane for $k_z=-0.1$ and $k_{||}$ changes from $-\pi$ to $\pi$, from which one can see $\nu = 1$. The winding number $\nu$ as a function of $k_z$ is shown in Fig. \ref{fig3} (d),
and compared with Fig. \ref{fig3}(b), one can see the zero-energy flat band appears in the $k_z$ momentum regime
where $\nu$ is non-zero. Thus, our results demonstrate the existence of mirror symmetry protected nontrivial
topological superconducting phase with flat zero-energy Majorana surface bands for the case of $\tilde{\Delta^{(+)}}$ pairing.
By following a similar method, we find that $\xi_{1,2,3,6,9}$ only lead to the trivial topological phase while $\xi_{4,5,7,8}$ are possible to give the nontrivial topological phases.

\begin{table}[htb]
  \centering
  \begin{minipage}[t]{1.\linewidth}
  \centering
	  \caption{Parameters of Kane model for configuration $\tilde{\mathcal{M}}_+$.}
\hspace{-1cm}
\begin{tabular}
[c]{ccccccccccccc}\hline\hline
 &$E_c$[eV]&$E_v$[eV]&P[$eV \cdot$ \AA]& $\gamma_1$&$\gamma_2$&$\gamma_3$&$V_{str}$[eV]&F&$\xi_7$\\
 $\tilde{\mathcal{M}}_+$&-1&0&8.46 & 4.1 &0.5&1.3&0&0&1 \\\hline\hline
\end{tabular}
\label{para_kane1}
  \end{minipage}
\end{table}

\subsubsection{Odd mirror parity case $\tilde{\mathcal{M}}_-$}
\label{oddMirrorParity}
Due to the block digonal nature of the BdG Hamiltonian, we can again only focus on the
block in the mirror parity $+i$ subspace with the gap function given by
\begin{eqnarray}
&&\Delta^{(-)}_{+i} = \left( 	
\begin{array}{ccc}
0&\eta_1+i\eta_2&i\eta_5+\eta_6\\
-(\eta_1+i\eta_2)&0&-\eta_3+i\eta_4\\
-(i\eta_5+\eta_6)&\eta_3-i\eta_4&0
\end{array}
\label{chapter4-gap-m-}
\right)
\end{eqnarray}
on the basis set by $\psi^{(-)} = (c_{+i}(\bold{k}), c^\dag_{+i}(-\bold{k}))^T$, is
The resulting Hamiltonian in the mirror parity $+i$ subspace is expressed as
\begin{eqnarray}
&&H^{(-)}_{+i} = \left( 	
\begin{array}{cc}
e_{+i}(\bold{k})&\Delta^{(-)}_{+i}\\
h.c.&-e^*_{+i}(-\bold{k})
\end{array}
\right)
\end{eqnarray}
where $e_{+i}$ is $3\times 3$ matrices, as defined in Eq. \ref{eq:+i}. This Hamiltonian respects particle-hole symmetry $C^{(-)}_{+i} = \tau_x \otimes \mathcal{I}_{3 \times 3}K$, and thus allows for the TSC phase
defined by Chern number as discussed above.

\begin{figure}[tb]
    \includegraphics[width = 1\columnwidth,angle=0]{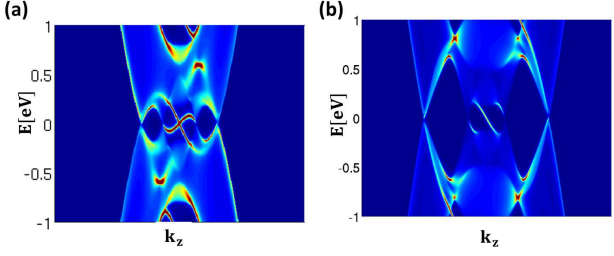}
  \caption{(Color online) (a) Surface DOS in mirror parity $+i$ subspace for $\tilde{M}_-$ configuration with two chiral edge modes. The chemical potential is $\mu = -0.5$ eV and the pairing magnitude $\eta_5 = 1.5$. (b) Surface DOS in mirror parity $+i$ subspace for $\tilde{M}_-$ configuration with one chiral edge mode. The chemical potential is $\mu = -0.5$ eV and the pairing magnitude $\eta_5 = 3$.}
    \label{fig4}
\end{figure}

Here we take $\eta_5$ term in the $T_1$ representation in Eq. (\ref{chapter4-gap-m-}) as an example and discuss other pairing functions later. For a small $\eta_5=1.5$ and other parameters listed in Table \ref{para_kane2}, the surface LDOS for a semi-infinite system along $k_\parallel$ direction is shown in Fig. \ref{fig4} (a), where two chiral Majorana surface modes exist in the mirror subspace +i, corresponding to the mirror TSC phase with the mirror Chern number $n_{+i} = -2$\cite{teo2008,zhangfan2013}.
When increasing the pairing function magnitude, a topological phase transition can occur and drive the system into the TSC phase
with only one chiral Majorana edge mode in the mirror $+i$ subspace.
Fig. \ref{fig4} (b) shows the LDOS at the surface for $\eta_5=3$,
which corresponds to the mirror TSC phase with mirror Chern number $n_M = -1$. We find the above result is quite general once a full bulk superconducting gap is opened in this case (Otherwise it will be nodal superconductivity).
This is because two by two block in the Hamiltonian $e_{+i}(\bold{k})$ spanned by the basis $\Phi_{+i,1}$ and
$\Phi_{+i,2}$ carries non-zero Chern number, which accounts for topological
surface state between the $\Gamma_6$ and $\Gamma_8$ bands. As a result, a mirror Chern number $n_{+i} = -2$
has been ``hidden'' in this system even without superconducting gap. Similar situation has been
discussed in the quantum anomalous Hall insulator in proximity to superconductivity\cite{qi2010chiral}.


\begin{table}[htb]
  \centering
  \begin{minipage}[t]{1.\linewidth}
  \centering
	  \caption{Parameters of Kane model for configuration $\tilde{\mathcal{M}}_-$.}
\hspace{-1cm}
\begin{tabular}
[c]{ccccccccccccc}\hline\hline
 &$E_c$[eV]&$E_v$[eV]&P[$eV \cdot$ \AA]& $\gamma_1$&$\gamma_2$&$\gamma_3$&$V_{str}$[eV]\\
$\tilde{\mathcal{M}}_-$ &-1&0&8.46 & 4.1 &0.5&1.3&1 \\\hline\hline
\end{tabular}
\label{para_kane2}
  \end{minipage}
\end{table}

\section{Topological surface superconductivity in half-Heusler compounds}
\label{projectionTSS}
In the above section, we have perform a systematic study of possible TSC phase in superconducting half-Heusler compounds based on the classification of bulk superconducting gap functions.
A recent scanning tunneling microscopy (STM) measurement of the superconducting gap \cite{banerjee2015}
suggests that surface superconductivity in LuPtBi
with superconducting transition temperature $T_c \approx 6 - 7$ K,
much greater than bulk transition temperature $T_c = 0.9K$ measured from transport experiment.
This motives us to study superconductivity related to surface states in this section,
rather than the bulk superconductivity.

\begin{widetext}
\begin{center}
\begin{table}[htb]
  \centering
  \begin{minipage}[t]{1.\linewidth}
  \centering
	  \caption{Projected gap functions on TSSs to the leading order of $k$ with $k=\sqrt{k_x^2+k_y^2}$ and $k_{\pm}=k_x\pm k_y$.}
\label{gapdecomposeSur}
\hspace{-1cm}
\begin{tabular}
[c]{c|c|c|c}\hline\hline
$\delta_i$     &Irrep  & Projection on Fermi surface  & Projection on two band model  \\\hline
$\delta_s$ &$A_1$[$\Gamma_1$]&$i e^{i\theta_k}$ & $\left( 	
\begin{array}{cc}
&1\\
-1&
\end{array}
\right)$\\ \arrayrulecolor{green}\hline\arrayrulecolor{black}
$\delta_1$ &$A_1$[$\Gamma_1$]&$i e^{i\theta_k}$ & $\left( 	
\begin{array}{cc}
&1\\
-1&
\end{array}
\right)$ \\
\hline
\arrayrulecolor{black}\hline\hline
$\delta_2$ & &$i e^{i\theta_k}$ & $\left( 	
\begin{array}{cc}
&1\\
-1&
\end{array}
\right)$\\
$\delta_4$ & E[$\Gamma_3$]&$i k(e^{-i\theta_k} + e^{i3\theta_k})$ & $i\left( 	
\begin{array}{cc}
k_+&\\
&k_-
\end{array}
\right)$\\\arrayrulecolor{green}\hline\arrayrulecolor{black}
$\delta_{11}$ &&0 & $\left( 	
\begin{array}{cc}
0&0\\
0&0
\end{array}
\right)$ \\
$\delta_{13}$ & E[$\Gamma_3$]&$k(e^{-i\theta_k} -e^{i3\theta_k})$ &  $\left( 	
\begin{array}{cc}
k_+&\\
&-k_-
\end{array}
\right)$\\ \arrayrulecolor{black}\hline\hline
$\delta_3$  &&$k(e^{-i\theta_k} -e^{i3\theta_k})$ &  $\left( 	
\begin{array}{cc}
k_+&\\
&-k_-
\end{array}
\right)$\\
$\delta_5$& &0 & $\left( 	
\begin{array}{cc}
&k_y\\
k_y&
\end{array}
\right)$\\
$\delta_6$  & $T_2$[$\Gamma_4$]&0 & $\left( 	
\begin{array}{cc}
&k_x\\
k_x&
\end{array}
\right)$\\\arrayrulecolor{green}\hline\arrayrulecolor{black}

$\delta_7$ &&0 & $\left( 	
\begin{array}{cc}
&k_y\\
k_y&
\end{array}
\right)$\\
$\delta_8$&&0 & $\left( 	
\begin{array}{cc}
&k_x\\
k_x&
\end{array}
\right)$\\
$\delta_{12}$&$T_2$[$\Gamma_4$]&$i e^{i\theta_k}$ &  $\left( 	
\begin{array}{cc}
0&1\\
-1&0
\end{array}
\right)$\\
\arrayrulecolor{black}\hline\hline

$\delta_9$ &&0 & $\left( 	
\begin{array}{cc}
&k_y\\
k_y&
\end{array}
\right)$\\
$\delta_{10}$ &&0 &  $\left( 	
\begin{array}{cc}
&k_x\\
k_x&
\end{array}
\right)$\\
$\delta_{14}$ &$T_1$[$\Gamma_5$]&$i k(e^{-i\theta_k} +e^{i3\theta_k})$ & $i\left( 	
\begin{array}{cc}
k_+&\\
&k_-
\end{array}
\right)$ \\
\hline\hline
$\delta_{d}$ & $A_1$[$\Gamma_1$]& $i e^{i\theta_k}$ &
$\left( 	
\begin{array}{cc}
0& 1\\
-1 & 0
\end{array}
\right)
$
\\
 \arrayrulecolor{green}\hline\arrayrulecolor{black}
$\delta_p$ & $A_1$[$\Gamma_1$]&$k (e^{-i\theta_k} - e^{i3\theta_k})$ &
$\left( 	
\begin{array}{cc}
 k_+&  0\\
0 & - k_-
\end{array}
\right)
$
 \\
\hline\hline
\end{tabular}
  \end{minipage}
\end{table}
\end{center}
\end{widetext}

\subsection{Projection of gap functions onto the Fermi surface of surface states}
In this section, our study starts with projecting bulk gap functions onto the TSSs of half-Heusler compounds numerically by applying $\tilde{\Delta}_s(\bold{k}) = \phi_s^\dag(\bold{k}) \Delta \phi_s^*(-\bold{k})$, where $\phi_s(\bold{k})$ is the eigen-wavefunction of TSSs at the momentum ${\bf k}$ and can be obtained by constructing a slab model with z direction as the open boundary. The relationship between the bulk gap functions and the projected surface gap functions is listed in Table \ref{gapdecomposeSur}.
For s-wave on-site pairing, three types of projected functions, including $p_x + ip_y$ [or $e^{i\theta_k}$], $(p_x - ip_y) + (p_x + ip_y)^3$ [or $e^{-i\theta_k} + e^{i3\theta_k}$] and $(p_x - ip_y) - (p_x + ip_y)^3$ [or $e^{-i\theta_k} - e^{i3\theta_k}$], can be obtained, where $\theta_k = tan^{-1}(\frac{k_y}{k_x})$.
In particular, $\delta_{s,1,2,12}$ lead to $p_x + ip_y$ pairing, $\delta_{4,14}$ give rise to
the $(p_x - ip_y) + (p_x + ip_y)^3$ pairing while $\delta_{3,13}$ yield $(p_x - ip_y) - (p_x + ip_y)^3$ pairing. For the pairings $\delta_{5,6,7,8,9,10,11}$, the projected gap function on the Fermi surface is exactly zero, indicating that nodal lines can be induced for TSSs.
The p-wave pairing $\delta_p$ is projected to $(p_x - ip_y) - (p_x + ip_y)^3$ pairing while the d-wave pairing $\delta_{d}$ corresponds to $p_x + ip_y$ pairings to the leading order of $k=\sqrt{k_x^2+k_y^2}$.

The above numerical results can also be extracted from analytical calculation of projecting gap functions onto TSSs,
for which the eigen-wavefunction of the upper half Dirac cone can be solved as $\phi_s(\mathbf{k}) =( f(z)/\sqrt{2})(aie^{-i\theta_k},a,bie^{-i\theta_k},b,-cie^{-i2\theta_k},ce^{i\theta_k})^T$ on the basis  $\Psi_e = (\Psi_{\Gamma_{6}},\Psi_{\Gamma_{8}})^T$,
where $(|a|^2 + |b|^2 + |c|^2) = 1$, $a,b=O(1)$, $c=O(k)$, $f(z)$ is the z-direction wavefunciton. Moreover, due to time-reveral symmetry, we can choose $a,c$ to be real and $b,f(z)$ to be imaginary. Let us take the pairing function $\delta_4 = -c_{\Gamma_{81},\uparrow}c_{\Gamma_{83},\uparrow} +c_{\Gamma_{81},\downarrow}c_{\Gamma_{83},\downarrow}$ as an example. By projecting the pairing functions into the TSS wavefunction, one can obtain $\tilde{\Delta}_4 =  \phi_s^\dag(\bold{k}) \delta_4 \phi_s^*(-\bold{k}) = F_zb^*c^*(e^{-i\theta_k} + e^{i3\theta_k})$
where $F_z = \sum_z f^*(z)f^*(z)$. This verifies the numerical results of the projection of gap function $\delta_4$.
Similar calculations can be applied to other gap functions to confirm the results in Table \ref{gapdecomposeSur}.

\subsection{Projection of gap functions onto two band effective model of surface states }
It is known that TSS of topological insulators can be described by the two band
Dirac Hamiltonian $H_{s,0}=A_0(k_y\sigma_x-k_x\sigma_y)-\mu$ on the basis $|\uparrow\rangle$ and
$|\downarrow\rangle$ which at ${\bf k}=0$ become the eigen-wavefunctions of TSSs.
Thus, we next try to construct projected gap functions on the basis $|\uparrow\rangle$ and
$|\downarrow\rangle$. Generally, we can write down an effective surface BdG Hamiltonian as
\begin{widetext}
\begin{eqnarray}
H_{s,BdG}=\left(
	\begin{array}{cccc}
          -\mu&A_0(k_y + ik_x)&\Delta_a&\Delta_b\\
          A_0(k_y - ik_x)&-\mu&\Delta_c&\Delta_d\\
          &&\mu&A_0(k_y - ik_x)\\
          h.c&&A_0(k_y + ik_x)&\mu
    \end{array}
	\right)
\label{eq:BdG_surface}
\end{eqnarray}
\end{widetext}
on the basis set by the operators $ (c_{\uparrow}(\mathbf{k}),c_{\downarrow}(\mathbf{k}),c^\dag_{\uparrow}(-\mathbf{k}),c^\dag_{\downarrow}(-\mathbf{k}))$,
where $c_{\uparrow,\downarrow}$ are the annihilation operators for $|\uparrow\rangle$ and
$|\downarrow\rangle$. It should be pointed out that the effective Hamiltonian of two band model $H_{s,0}$
for surface states possesses quite high symmetry, including full rotation symmetry and in-plane mirror
symmetry along any direction.

To get the form of $\Delta$ function in the two band model
for surface states, one can write down the explicit form of basis wave functions of surface states,
given by $|\uparrow\rangle=f(z)(a,0,b,0,-ce^{-i\theta_k},0)^T$ and $|\downarrow\rangle=f(z)
(0,a,0,b,0,ce^{i\theta_k})^T$, and project the gap function $\delta_i$ onto these two basis wave functions
directly. The obtained gap functions (2 by 2 matrices) are shown in the third column in Table \ref{gapdecomposeSur}.
We can further project gap functions on the Fermi surface of surface states with the eigen wavefunction
given by $\phi_s(\mathbf{k}) =\frac{1}{\sqrt{2}} (ie^{-i\theta_k},1)^T$ on the basis $|\uparrow\rangle$ and $|\downarrow\rangle$. As a result, the projection of gap functions
can be obtained as $\tilde{\Delta} =  \phi_s^\dag(\bold{k}) \Delta \phi_s^*(-\bold{k})=
\frac{1}{2}( \Delta_a e^{i2\theta_k} + \Delta_d) +  \frac{1}{2}ie^{i\theta_k}( \Delta_c - \Delta_b)$,
which are consistent with the results obtained in the last section (the second column in
Table \ref{gapdecomposeSur}).

Next, we will discuss the band dispersion of the surface BdG Hamiltonian using the result in the third column in Table \ref{gapdecomposeSur}. For simplicity, we choose the overall phase of the gap function in the third column of Table \ref{gapdecomposeSur} to be real, thus consistent with the TR symmetry defined by $\Theta = i\tau_0\sigma_y K$ on the BdG Hamiltonian Eq. (\ref{eq:BdG_surface}). Below we will discuss different pairing forms separately.


The $\tilde{\Delta}=p_x+ip_y$ pairing function corresponds to the choice of  $\Delta_b = - \Delta_c = \Delta_A$ and $\Delta_a = \Delta_d = 0$, which gives rise to $\tilde{\Delta} = -i \Delta_A e^{i\theta_k} = -i\Delta_A \frac{k_x +ik_y}{|k|}$. The eigenenergy of the corresponding Hamiltonian can be solved as $E_A = \pm \sqrt{(A_0k\pm \mu)^2 + \Delta_A^2 }$,
which gives a fully gaped superconductivity on TSS. This situation occurs for the gap functions
$\delta_{s,1,2,12}$, corresponding to $\xi_{1,2,3,4}$ according to Table \ref{correspondence} in Appendix \ref{correspondence_delta_xi}, as well as $\delta_{d}$.
It should be mentioned that although the energy dispersion is gaped for the isotropic d-wave quintet pairing function $\delta_{d}$ to the leading order of $k$, the nodal points/lines can exist if including terms with high-order $k$. For example, if including terms of $k^2$, we have $\Delta_b=-\Delta_c=\Delta_D (1+\alpha k^2)$ and $\Delta_a=\Delta_d=0$, and the system has a nodal circle for $\alpha=-A_0^2/\mu^2$.

For the $\tilde{\Delta}=(p_x - ip_y) +(p_x+ip_y)^3 $ pairing, one may choose $\Delta_a = i\Delta_B(k_x + ik_y)$, $\Delta_d = i\Delta_B(k_x - ik_y)$ and $\Delta_b = \Delta_c = 0$, which give rise to
$\tilde{\Delta} = \frac{i}{2}( \Delta_B k e^{i3\theta_k} + \Delta_Bk e^{-i\theta_k})$.
The corresponding eigenenergy is
$E_B = \pm \sqrt{A_0^2k^2 + \mu^2+\Delta_B^2k^2 \pm 2 |A_0|\sqrt{\mu^2 k^2 + 4\Delta_B^2k_x^2k_y^2}}$
with four nodes $(k_x,k_y) = (\pm \sqrt{\frac{\mu^2}{2(A_0^2 - \Delta_B^2)}}, \pm \sqrt{\frac{\mu^2}{2(A_0^2 - \Delta_B^2)}})$ along $\hat{k}_\perp$ and $\hat{k}_\parallel$ directions in the momentum space with $\Delta_B^2 < A_0^2$. We notice that the momentum lines $\hat{k}_\perp=0$ and $\hat{k}_\parallel=0$ respect
the mirror symmetry $\tilde{M}_{[110]} = diag[M_{[110]}, -M^*_{[110]}]$ with $M_{[110]} = e^{i\frac{(s_x+s_y)\pi}{\sqrt{2}}}$ and these nodal points on the surface are protected by mirror symmetry,
which indicates non-trivial bulk topology in the corresponding mirror invariant plane in 3D bulk.

\begin{figure}[tb]
    \includegraphics[width = 0.9\columnwidth,angle=0]{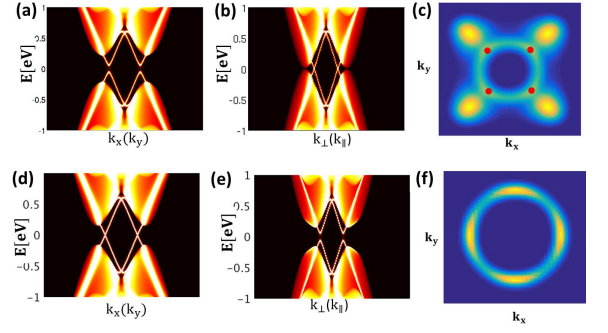}
  \caption{(Color online) (a-c) Energy dispersion for gap function $\delta_4$ along $\hat{k}_{x,y}(\hat{k}_{\perp,\parallel})$ directions with z as its open boundary condition (a, b) and the corresponding  surface DOS in the momentum space(c), which corresponds to the case $(p_x - ip_y) + (p_x+ip_y)^3$. (d-f) Energy dispersion for gap function $\delta_3$ along $\hat{k}_{x,y}(\hat{k}_{\perp,\parallel})$ directions with z as its open boundary condition (d, e) and the corresponding  surface DOS in the momentum space(f), which corresponds to the case $(p_x - ip_y) - (p_x+ip_y)^3$.}
    \label{fig4c}
\end{figure}

This situation occurs for the pairing functions $\delta_{4,14}$, corresponding to $\eta_{3,5}$
used in Sec. \ref{oddMirrorParity}, according to Table \ref{correspondence} in Appendix \ref{correspondence_delta_xi}. 
We also perform a slab model calculation with z as its open boundary condition for the case with gap function $\delta_4$. The parameters are listed in Table \ref{para_kane2} with the chemical potential $\mu = -0.1$ eV and gap function magnitude $\eta_3 = 0.3$. The energy dispersion along $\hat{k}_{x,y}(\hat{k}_{\perp,\parallel})$ directions are shown in panel (a-b) in Fig. \ref{fig4c}. We notice that there are two helical modes along [110] directions, which are consistent with the case of $\eta_5$ we discussed in Sec. \ref{oddMirrorParity} with the mirror Chern number $-2$. The corresponding surface DOS is also illustrated in Fig. \ref{fig4c}.
Four surface Dirac points are located on the inner ring, as depicted by four red dots in Fig. \ref{fig4c}c, while four dark yellow regions outside the ring are caused by enhanced bulk DOS.



For surface gap function $\tilde{\Delta}=(p_x - ip_y) - (p_x+ip_y)^3$,  one possible choice is $\Delta_a = \Delta_C(k_x + ik_y)$, $\Delta_d = -\Delta_C(k_x - ik_y)$ and $\Delta_b = \Delta_c = 0$, which results in the projected gap function on the upper Dirac cone as $\tilde{\Delta} = \frac{1}{2}( \Delta_C k e^{i3\theta_k} - \Delta_C k e^{-i\theta_k})$. Similarly, one can solve for the eigen energy of the BdG Hamiltonian as $E_C = \pm \sqrt{A_0^2k^2 + \mu^2+\Delta_C^2k^2 \pm 2 |A_0|\sqrt{\Delta_C^2(k_x^2 - k_y^2)^2 + \mu^2 k^2}}$. Thus, the system owns four total nodal points $(k_x,k_y) = (\pm \sqrt{\frac{\mu^2}{A_0^2 - \Delta_C^2}}, 0), (0, \pm \sqrt{\frac{\mu^2}{A_0^2 - \Delta_C^2}})$ along $\hat{k}_x$ and $\hat{k}_y$ directions. One correspondence of this case is the situation with gap function $\delta_3$ in the $\tilde{M}_+$ configuration. However, the surface nodal points in this case is not the same as the discussion of the AIII symmetry class of $\delta_3$ (or the corresponding $\xi_6$ according to the table IX) on the $\mathcal{M}_{(110)}$ mirror plane. Instead, the nodal points exist in the $k_x$ and $k_y$ lines due to the additional mirror symmetry
with respect to (100) plane in the Kane model.
We find that the original BdG Hamiltonian in each mirror parity subspace belongs to symmetry class D for $\delta_3$ on mirror invariant planes (100), as shown in Appendix \ref{D_for_delta3}, where chiral edge states can exist in each mirror parity subspace. Panel (d-e) in Fig. \ref{fig4c} shows energy dispersion along $\hat{k}_{x,y}(\hat{k}_{\perp,\parallel})$ directions for the case with gap function $\delta_3$ calculated from a slab model with open boundary condition along z direction. The parameters are lised in Table \ref{para_kane2} with the chemical potential $\mu = -0.1$ eV and gap function magnitude $\xi_6 = 0.3$. The helical modes along $\hat{k}_{x,y}$ directions result in four Dirac surface nodes, as shown in Fig. \ref{fig4c} (c). Thus, the four nodal points on mirror invariant planes (100) for the surface BdG Hamiltonian also come from the mirror symmetry protected topological phase.

Finally, we would like to mention the behaviors of $\delta_{5,6,7,8,9,10}$ with zero gap functions when projecting onto the Fermi surface of TSSs. However, the corresponding projected gap functions for the two band models are non-zero and shown in the third column of Table \ref{gapdecomposeSur}. By choosing $\Delta_a=\Delta_d=\Delta_x k_x+\Delta_y k_y$ (with $\Delta_x=0$ for $\delta_{5,7,9}$ and $\Delta_y=0$ for $\delta_{6,8,10}$) and $\Delta_c=\Delta_b=0$, the eigenenergy become $E_D=\pm||A_0 k|\pm\sqrt{(\Delta_x k_x+\Delta_y k_y)^2+\mu^2}|$. Therefore, nodal line exists when $A_0^2 k_x^2+(A_0^2-\Delta_y^2)k_y^2=\mu^2$ for $\delta_{5,7,9}$ or $(A_0^2-\Delta_x^2) k_x^2+A_0^2 k_y^2=\mu^2$ for $\delta_{6,8,10}$. The existence of nodal line is consistent with the zero gap functions when projecting on the Fermi surface.

\subsection{Domain wall states between $p_x + ip_y$ and $(p_x - ip_y) \pm(p_x+ip_y)^3$ domains}
One interesting consequence between two topological distinct phases is the existence of topological protected zero modes at the interface. Here, the interface states between surface domains with $p_x + ip_y$ and $(p_x - ip_y) \pm(p_x+ip_y)^3$ pairing states will be explored by constructing a periodic superlattice with alternating $p_x + ip_y$ and $[(p_x - ip_y) \pm (p_x+ip_y)^3]$ domains growing in the x direction with width L\cite{beugeling2012}. Surface superconductivity with $p_x + ip_y$ pairing lies in the region $[-\frac{L}{2},0]$ and Superconductivity with $(p_x - ip_y) \pm(p_x+ip_y)^3$ pairing lies in the region $[0,\frac{L}{2}]$. Due to the periodic boundary condition and the Bloch's theorem, the wavefunction can be written as
\begin{eqnarray}
\Psi_\xi = \frac{1}{2\pi}e^{i(k_x x + k_y y)} \vert U^\xi(x)_\bold{k}\rangle
\end{eqnarray}
Here $ U^\xi(x+L)_\bold{k}\rangle = U^\xi(x)_\bold{k}\rangle$ is a periodic function, which can be expanded in terms of plane-wave functions
\begin{eqnarray}
 \vert U^\xi(x)_\bold{k}\rangle = \sum_{n,\lambda} a^{\xi}_{n,\lambda} \vert n,\lambda \rangle = \sum_{n,\lambda} a^{\xi}_{n,\lambda} \frac{1}{\sqrt{L}} e^{i(2\pi n/L)x}\vert \lambda \rangle
\end{eqnarray}
where $\vert \lambda \rangle$ labels the component $\lambda = 1, 2, 3, 4$ of the wave function. Assume that $H \Psi_\xi = E_\xi \Psi_\xi$. By expanding the wavefunction in terms of plane-wave function, one arrives at
\begin{eqnarray}
\sum_{n',\lambda'} \langle n,\lambda \vert \hat{H} \vert n',\lambda' \rangle a^{\xi}_{n',\lambda'} = E_\xi a^{\xi}_{n,\lambda}
\end{eqnarray}

\begin{figure}[tb]
    \includegraphics[width = 0.9\columnwidth,angle=0]{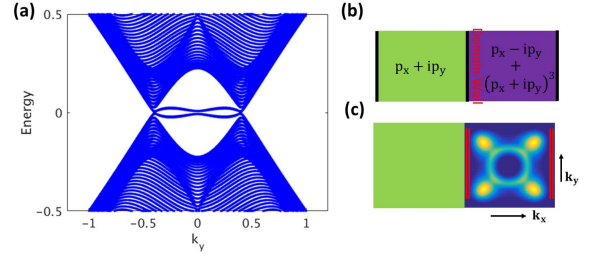}
  \caption{(Color online) (a) Bulk band structure of constructed superlattice for $p_x + ip_y/(p_x - ip_y) +(p_x+ip_y)^3$ case. (b) Geometric structure of the constructed superlattice. (c) Schematic plot of surface DOS in the momentum space. The two red lines at each domain wall show the projected doubly boundary modes. The parameters we are using are $A_0 =1$, $\Delta_B = 0.5$, $\mu = 0.5$, $\Delta_A  = 0.5$, $L = 2000$ and $N = 50$ .}
    \label{fig5}
\end{figure}

Since the low energy physics plays the essential role, one may only take a finite number of n states, denoted as $-N, -N+1, ..., N-1, N$ with $N = 50$. The band structure for the constructed superlattice is shown in Fig. \ref{fig5} for case of $p_x + ip_y/(p_x - ip_y) + (p_x+ip_y)^3$ configuration. The geometric structure of the superlattice is shown in Fig. \ref{fig5} (b). The surface DOS in the momentum space for each domain is shown in Fig. \ref{fig5} (c) and four nodal points in the momentum space for the domain $(p_x - ip_y) + (p_x+ip_y)^3$ emerge along $k_x = k_y \neq 0$ momentum lines on the mirror invariant planes. It is found that there exist four modes around zero energy at domain walls, as illustrated in Fig. \ref{fig5} (a). The four interface modes come from two copies of doubly degenerate boundary modes between  $p_x + ip_y/(p_x - ip_y) + (p_x+ip_y)^3$ domains, which originate from the projection of states onto the domain walls due to the four nodal points in the momentum space in the domain $(p_x - ip_y) + (p_x+ip_y)^3$. The doubly degenerate interface modes interact with each other and lift the zero-energy degeneracy, as shown in Fig. \ref{fig5}(a).

\begin{figure}[tb]
    \includegraphics[width = 0.9\columnwidth,angle=0]{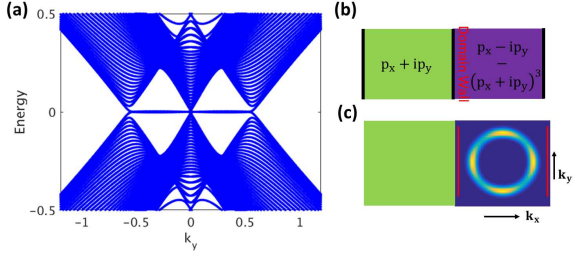}
  \caption{(Color online) (a) Bulk band structure of constructed superlattice for $p_x + ip_y/(p_x - ip_y) -(p_x+ip_y)^3 $ case. (b) Geometric structure of the constructed superlattice. (c) Schematic plot of surface DOS in the momentum space. The red line at each domain wall shows the projected singly boundary mode. The parameters we are using are $A_0 =1$, $\Delta_C = 0.5$, $\mu = 0.5$, $\Delta_A  = 0.5$, $L = 2000$ and $N = 50$ .}
    \label{fig6}
\end{figure}

Similarly, one can calculate the band structure for the constructed superlattice for case of $p_x + ip_y/(p_x - ip_y) - (p_x+ip_y)^3$, as shown in Fig. \ref{fig6} (a).  One can see that there exist two-fold degenerate zero-energy modes, which indicates the emergence of nontrivial topological phases. The robustness of the interface zero-energy modes originates the singly projected boundary states in the $(p_x - ip_y) - (p_x+ip_y)^3$ domains, where the four nodal points in the momentum space lie along the $k_x = 0$ and $k_y = 0$ momentum lines, as illustrated in Fig. \ref{fig6}. The essential difference between the $p_x + ip_y/(p_x - ip_y) + (p_x+ip_y)^3$ case and the $p_x + ip_y/(p_x - ip_y) - (p_x+ip_y)^3$ case is the positions of the nodal points of domains $(p_x - ip_y) \pm (p_x+ip_y)^3$ in the momentum space since the flat Majorana bands come from the projection of the nodal points onto the domain wall, as shown in Fig. \ref{fig5} (c) and Fig. \ref{fig6} (c).

\section{Discussion and conclusion}
In conclusion, we have presented a systematical study of possible s-wave gap functions allowed in half-Heusler compounds. We find that the on-site pairing states with higher angular momenta could also be present when projecting the pairing function into the band near the Fermi energy and lead to nodal superconducting phases
\cite{PhysRevB.85.024522,PhysRevB.84.020501,PhysRevB.83.064505,schnyder2011}
.
Furthermore, the mirror symmetry plays an essential role in the topological classification for half-Heusler superconductors. The model reveals that the BdG Hamiltonian on the mirror invariant planes can be in the symmetry class AIII for even mirror parity pairing functions,  or symmetry class D for odd mirror parity pairing functions. Moreover, by projecting gap functions into TSSs, three types of gap functions, including $p_x + ip_y$, $(p_x - ip_y) + (p_x + ip_y)^3$ and $(p_x - ip_y) - (p_x + ip_y)^3$ are identified. The domain wall between the $p_x+ip_y$ and $(p_x - ip_y)\pm (p_x+ip_y)^3 $ pairing states can host possible zero-energy flat bands.
It should be emphasized that our results neglect the influence from the inversion symmetry breaking term
(C term in the $\Gamma_8$ part of Kane model). Involving the inversion symmetry breaking term will gap out some nodal lines and nodal rings found in the Sec. \ref{projectionTSS}  
for TSSs. However, such inversion symmetry breaking
is quite small compared to other energy scale in the 6 band Kane model. Particularly, We expect such term only appears in higher order of angular momenta for two band model of surface states.
Thus, these nodal points and nodal rings should remain within certain energy regime.
Our studies suggest the possibility of mirror TSC in superconducting half-Heusler compounds, including LnPtBi (Ln = Y, Lu)\cite{liuz2016, butch2011, tafti2013} and RPdBi (R = rare earth)\cite{Nakajimae1500242}.
Finally, we would like to describe briefly the future direction from this work. The current work only involves the classification of gap function in the Kane model, but has not considered the Ginzburg-Landau free energy or equivalently the self-consistent gap equation. Thus, the question which pairing will be energetically favored in the Kane model has not been answered in this work. We notice several recent works have suggested the possibility of mixed-paring states in half-Heusler compounds, including the s-wave singlet and p-wave septet mixing\cite{brydon2016,PhysRevB.96.144514}  and s-wave singlet and d-wave quintet mixing\cite{yu2017singlet}.
Our work here only focused on a single type of pairing state and has not taken into account mixed-pairing states. It will be an interesting question how these mixed-pairing states are projected into the surface energy spectrum. Moreover, how to distinguish different types of pairing states in experiments will be another important question for the future work.

\bibliography{unconventionalSC}

\begin{appendix}
\begin{widetext}
\section{Kane model}
The eight basis functions in Kane model\cite{novik2005,winkler2003} are
\begin{eqnarray}
\nonumber&&\vert \Gamma_6,1/2 \rangle =  \vert S \rangle \vert \uparrow \rangle \\
\nonumber&&\vert \Gamma_6,-1/2 \rangle = \vert S \rangle \vert \downarrow \rangle \\
\nonumber&&\vert \Gamma_8,1/2 \rangle =  \frac{1}{\sqrt{6}}(2 \vert Z \rangle \vert \uparrow \rangle - \vert  X + iY \rangle \vert \downarrow \rangle)\\
\nonumber&&\vert \Gamma_8,-1/2 \rangle =  \frac{1}{\sqrt{6}}(2 \vert Z \rangle \vert \downarrow \rangle + \vert  X - iY \rangle \vert \uparrow \rangle)\\
\nonumber&&\vert \Gamma_8,3/2 \rangle = -\frac{1}{\sqrt{2}}\vert  X + iY \rangle \vert \uparrow \rangle\\
\nonumber&&\vert \Gamma_8,-3/2 \rangle = \frac{1}{\sqrt{2}}\vert  X - iY \rangle \vert \downarrow \rangle\\
\nonumber&& \vert \Gamma_7,1/2 \rangle = -\frac{1}{\sqrt{3}} (\vert  Z \rangle \vert \uparrow \rangle + \vert X +iY \rangle \vert \downarrow \rangle)\\
&& \vert \Gamma_7,-1/2 \rangle = \frac{1}{\sqrt{3}} ( \vert  Z \rangle \vert \downarrow \rangle - \vert X - iY \rangle \vert \uparrow \rangle)
\label{eqn:basis}
\end{eqnarray}, where $\vert X \rangle$, $\vert Y \rangle$ and $\vert Z \rangle$ are real and $\vert S \rangle$ is purely imaginary.

On such a choice of basis functions, the Kane model is expressed as

\begin{eqnarray}
  &&H_{Kane}=\\
\nonumber  &&\left(
	\begin{array}{cccccccc}
          T&0&\sqrt{\frac{2}{3}}Pk_z& \frac{1}{\sqrt{6}}Pk_-& -\frac{1}{\sqrt{2}}Pk_+&0&-\frac{1}{\sqrt{3}}Pk_z&-\frac{1}{\sqrt{3}}Pk_-\\
          0&T&-\frac{1}{\sqrt{6}} Pk_+& \sqrt{\frac{2}{3}}Pk_z&0 & \frac{1}{\sqrt{2}}Pk_- &-\frac{1}{\sqrt{3}}Pk_+ & \frac{1}{\sqrt{3}}Pk_z\\
          \sqrt{\frac{2}{3}}Pk_z&-\frac{1}{\sqrt{6}} Pk_-&U-V&C&-\bar{S}^{\dag}_-&R&\sqrt{2}V&-\sqrt{\frac{3}{2}}\tilde{S}_-\\
	  \frac{1}{\sqrt{6}} Pk_+&\sqrt{\frac{2}{3}}Pk_z&C^{\dag}&U-V&R^{\dag}&\bar{S}^{\dag}_+&-\sqrt{\frac{3}{2}}\tilde{S}_+&-\sqrt{2}V\\
	  -\frac{1}{\sqrt{2}}Pk_-&0&-\bar{S}_-&R&U+V&0&\frac{1}{\sqrt{2}}\bar{S}_-&-\sqrt{2}R\\
	  0&\frac{1}{\sqrt{2}}Pk_+&R^{\dag}&\bar{S}_+&0&U+V&\sqrt{2}R^\dag&\frac{1}{\sqrt{2}}\bar{S}_+\\
        -\frac{1}{\sqrt{3}}Pk_z&-\frac{1}{\sqrt{3}}Pk_-&\sqrt{2}V&-\sqrt{\frac{3}{2}}\tilde{S}^\dag_+&\frac{1}{\sqrt{2}}\bar{S}^\dag_-&\sqrt{2}R&U-\Delta_{so}&C\\
        -\frac{1}{\sqrt{3}}Pk_+&\frac{1}{\sqrt{3}}Pk_z&-\sqrt{\frac{3}{2}}\tilde{S}^\dag_-&-\sqrt{2}V&-\sqrt{2}R^\dag&\frac{1}{\sqrt{2}}\bar{S}^\dag_+&C^\dag&U-\Delta_{so}
	\end{array}
\label{eq:ham_kane}
	\right)
\end{eqnarray}

where
\begin{eqnarray}
\nonumber&&T = E_c + \frac{\hbar^2}{2m_0}[(2F+1)k^2_{||}+k_z(2F+1)k_z]\ , \\
\nonumber&&U = E_v - \frac{\hbar^2}{2m_0}(\gamma_1 k^2_{||} + k_z \gamma_1 k_z)\ , \\
\nonumber&&V = - \frac{\hbar^2}{2m_0}(\gamma_2 k^2_{||} - 2 k_z \gamma_2 k_z)\ , \\
\nonumber&&R = - \frac{\hbar^2}{2m_0}(\sqrt{3} \mu k^2_+ - \sqrt{3} \bar{\gamma}k^2_-)\ , \\
\nonumber&&\bar{S}_{\pm} = - \frac{\hbar^2}{2m_0} \sqrt{3}k_{\pm}(\{\gamma_3,k_z\} + [\kappa,k_z])\ , \\
\nonumber&&\tilde{S}_\pm = - \frac{\hbar^2}{2m_0} \sqrt{3}k_{\pm}(\{\gamma_3,k_z\} -\frac{1}{3} [\kappa,k_z])\ , \\
\nonumber&&C =  \frac{\hbar^2}{2m_0}k_-[\kappa,k_z]\ , 
\end{eqnarray}
$k_{\parallel}^2=k_x^2+k_y^2$, $\bar{\gamma} = \frac{\gamma_2+\gamma_3}{2}$ and $\mu = \frac{-\gamma_2+\gamma_3}{2}$. $\{.\}$ and $[.]$ indicate the anticommutator and commutator. $\Delta_{so}$ is energy splitting caused by the spin-orbit coupling between $\Gamma_8$ and $\Gamma_7$ bands.

\section{Classification of Gap Function in $T_d$ Group}
$T_d$ point group can be generated by mirror symmetry operation with respect to (110) plane $M_{[110]}$, three-fold rotational operation along [111] direction $C_3$ and improper four-fold rotational operation along [001] direction $S_4$. Thus, we will use the three symmetry operations to find possible on-site Cooper pairings formed from states within $\Gamma_6$ band, states within $\Gamma_8$ band and states between $\Gamma_6$ and $\Gamma_8$ bands, and their corresponding IrReps.

\subsection{Character table of $T_d$ group and $T_d$ double group}
Before we perform the classification of gap function, let us take a look at the character table for $T_d$ group and $T_d$ double group, which are listed in Table \ref{tab1} and  \ref{tab2}. The first column lists all irreducible representations of $T_d$ and $T_d$ double group and the first row lists all possible symmetry operations in $T_d$ and $T_d$ double group. $\hat{E}$ in $T_d$ double group is an rotational operation of  $2\pi$ around a unit vector $\hat{n}$. $\hat{E}$ is equivalent to an identity operation for spin-$N$ systems with non-negative integer $N$, while $\hat{E}$ becomes negative identity for spin-$\frac{N}{2}$ systems with odd positive integer $N$.

\begin{center}
\begin{table}[htb]
  \begin{minipage}[t]{1\linewidth}
  \centering
      \caption{Character table of $T_d$ group.}
\label{tab1}
\hspace{0cm}
\begin{tabular}{ccccccc}
    \hline\hline
    & \{E\}\hfill & \{3$C_2$\} \hfill &\{6$S_4$\} \hfill & \{6$\sigma$ \} \hfill & \{8$C_3$\} \hfill &  Basis functions \\  \hline
  $A_1$&1 &1&1&1&1&$xyz$\\
  $A_2$&1 &1&-1&-1&1&$x^4(y^2-z^2)+y^4(z^2-x^2)+z^4(x^2-y^2)$\\
  $E$&2 &2&0&0&-1&$\{ (x^2-y^2), z^2 - \frac{1}{2}(x^2+y^2) \}$\\
  $T_1$&3 &-1&1&-1&0&$\{ x(y^2-z^2), y(z^2 - x^2), z(x^2-y^2)\}$\\
  $T_2$&3 &-1&-1&1&0&$\{ x, y, z\}$\\
   \hline \hline
  \end{tabular}
  \end{minipage}
\end{table}
\end{center}

\begin{center}
\begin{table}[htb]
  \begin{minipage}[t]{1\linewidth}
  \centering
      \caption{Character table of $T_d$ double group.}
\label{tab2}
\hspace{0cm}
\begin{tabular}{ccccccccc}
    \hline\hline
    & \{E\}\hfill & \{3$C_2/3\hat{E}C_2$\} \hfill &\{6$S_4$\} \hfill & \{6$\sigma/6\hat{E}\sigma$ \} \hfill & \{8$C_3$\} \hfill & $\hat{E}$ \hfill &\{6$\hat{E}S_4$\} \hfill &  \{ 8$\hat{E}C_3$ \} \\ \hline
  $\Gamma_1$&1 &1&1&1&1&1&1&1\\
  $\Gamma_2$&1 &1&-1&-1&1&1&-1&1\\
  $\Gamma_3$&2 &2&0&0&-1&2&0&-1\\
  $\Gamma_4$&3 &-1&-1&1&0&3&-1&0\\
  $\Gamma_5$&3 &-1&1&-1&0&3&1&0\\
  $\Gamma_6$&2 &0&$\sqrt{2}$&0&1&-2&-$\sqrt{2}$&-1\\
  $\Gamma_7$&2 &0&-$\sqrt{2}$&0&1&-2&$\sqrt{2}$&-1\\
  $\Gamma_8$&4 &0&0&0&-1&-4&0&1\\
   \hline \hline
  \end{tabular}
  \end{minipage}
\end{table}
\end{center}

\subsection{Cooper pairings from $\Gamma_6$ bands}
Now let us work out the three operators for $\Gamma_6$ bands. One would use the following matrices: $s_{0,i} = \frac{1}{2}\sigma_{0,i}$, where $s_0$ is two-by-two unit matrix and $\sigma_{i}$ are Pauli matrices with $i = \{x,y,z\}$. The rotational operator along $\hat{n}$ direction with angle $\theta$ is defined as $R_{\hat{n}}(\theta) = e^{i(\hat{n}\cdot \vec{s})\theta}$. Thus, the mirror reflection operator with respect to (110) plane for $\Gamma_6$ bands, denoted as $m_{[110],\Gamma_6}$, is
$m_{[110],\Gamma_6} = I C_2= Ie^{i(s_x+s_y)\pi/\sqrt{2}}$, where I is the inversion operator with I = 1 for $\Gamma_6$ bands. The $S_4$ operator $S_{4,\Gamma_6}$ is
$S_{4,\Gamma_6} = e^{is_z\pi/2} m_{z,\Gamma_6}$
where $m_{z,\Gamma_6} = I e^{is_z\pi}$. The $C_3$ operator $C_{3,\Gamma_6}$ is
$C_{3,\Gamma_6} = e^{i(s_x+s_y+s_z)2\pi/3\sqrt{3}}$. The s-wave gap function for $\Gamma_6$ band is $\delta_s = i\sigma_y$. One can check that
\begin{eqnarray}
\nonumber && m_{[110],\Gamma_6} \delta_s m^T_{[110],\Gamma_6} = \delta_s\\
\nonumber && S_{4,\Gamma_6} \delta_s S^T_{4,\Gamma_6} = \delta_s \\
&& C_{3,\Gamma_6} \delta_s C^T_{3,\Gamma_6} = \delta_s.
\end{eqnarray}
Thus, the on-site Cooper pair for $\Gamma_6$ bands ($-c_{\Gamma_{6},\uparrow}c_{\Gamma_{6},\downarrow} + c_{\Gamma_{6},\downarrow}c_{\Gamma_{6},\uparrow}$) belongs to $A_1$($\Gamma_1$) IrRep.

\subsection{Cooper pairings from $\Gamma_8$ bands}
\label{cooper_pairings_bands_gamma8}
In order to obtain the three operators for $\Gamma_8$ band, we define J matrices under the basis function $\Psi_{\Gamma_8} = (\vert \Gamma_8,1/2 \rangle,\vert \Gamma_8,-1/2 \rangle,\vert \Gamma_8,3/2 \rangle,\vert \Gamma_8,-3/2 \rangle)^T$, which are expressed as
$J_x=\frac{1}{2}\left(\begin{array}{cccc}
          0&2&\sqrt{3}&0\\
          2&0&0&\sqrt{3}\\
          \sqrt{3}&0&0&0\\
          0&\sqrt{3}&0&0
    \end{array}
	\right)$, $J_y=\frac{i}{2}\left(
	\begin{array}{cccc}
          0&-2&\sqrt{3}&0\\
          2&0&0&-\sqrt{3}\\
          -\sqrt{3}&0&0&0\\
          0&\sqrt{3}&0&0
    \end{array}
	\right)$ and $J_z=\frac{1}{2}\left(
	\begin{array}{cccc}
          1&0&0&0\\
          0&-1&0&0\\
          0&0&3&0\\
          0&0&0&-3
    \end{array}
	\right).$
The rotational operator along $\hat{n}$ direction with angle $\theta$ is defined as $R_{\hat{n}}(\theta) = e^{i(\hat{n}\cdot \vec{J})\theta}$. Thus, a two-fold rotational operator along [110] direction is defined as $R_{2,[110],\Gamma_8}(\pi) = e^{i(J_x+J_y)\pi/\sqrt{2}} $.
We define the inversion symmetry operator for $\Gamma_8$ bands as $I = -1$. Thus, the mirror symmetry along [110] direction is written as $m_{[110],\Gamma_8} = I C_{2,[110]}(\pi)$. The $S_4$ operator along [001] direction is written as $S_{4,\Gamma_8} = m_z R_z(\frac{\pi}{2})$, where $m_z$ is mirror operation along z direction. The $C_3$ rotation operator $C_{3,\Gamma_8}$ along [111] is $C_{3,\Gamma_8} = e^{\frac{i(Jx + Jy+Jz)2\pi}{3\sqrt{3}}}$.

With the help of the four-by-four antisymmetric matrices defined below, $\delta_1 =\tau_z\otimes i\sigma_y$, $\delta_2 =\tau_0\otimes i\sigma_y$, $\delta_3 =\tau_y\otimes \sigma_0$, $\delta_4 =i\tau_y\otimes \sigma_z$, $\delta_5 =i\tau_y\otimes \sigma_x$ and $\delta_6 =\tau_x\otimes \sigma_y$,  we are ready to do the gap function classification within $\Gamma_8$ band. One can check that the matrices above preserve TR symmetry and PH symmetry.

For $m_{[110],\Gamma_8}$ operation, we have
\begin{eqnarray}
\nonumber m_{[110],\Gamma_8} \delta_{1} m^T_{[110],\Gamma_8} &&= \delta_{1} \\
\nonumber ---\\
\nonumber m_{[110],\Gamma_8} \delta_{2} m^T_{[110],\Gamma_8} &&= \delta_{2} \\
\nonumber m_{[110],\Gamma_8} \delta_{4} m^T_{[110],\Gamma_8} &&= -\delta_{4} \\
\nonumber ---\\
\nonumber m_{[110],\Gamma_8} \delta_{3} m^T_{[110],\Gamma_8} &&= \delta_{3} \\
\nonumber m_{[110],\Gamma_8} \delta_{5} m^T_{[110],\Gamma_8} &&= \delta_{6} \\
\nonumber m_{[110],\Gamma_8} \delta_{6} m^T_{[110],\Gamma_8} &&= \delta_{5}
\end{eqnarray}

For $S_4$ operation, we have
\begin{eqnarray}
\nonumber S_4 \delta_{1} S^T_4 &&= \delta_{1} \\
\nonumber ---\\
\nonumber S_4 \delta_{2} S^T_4 &&= \delta_{2} \\
\nonumber S_4 \delta_{4} S^T_4 &&= -\delta_{4} \\
\nonumber ---\\
\nonumber S_4 \delta_{3} S^T_4 &&= -\delta_{3} \\
\nonumber S_4 \delta_{5} S^T_4 &&= -\delta_{6} \\
S_4 \delta_{6} S^T_4 &&= \delta_{5}
\end{eqnarray}

For $C_3$ operation, we have
\begin{eqnarray}
\nonumber C_3 \delta_{1} C^T_3 &&= \delta_{1} \\
\nonumber ---\\
\nonumber C_3 \delta_{2} C^T_3 &&= -\frac{1}{2}\delta_{2}  + \frac{\sqrt{3}}{2}\delta_{4}\\
\nonumber C_3 \delta_{4} C^T_3 &&=  -\frac{\sqrt{3}}{2}\delta_{2}  - \frac{1}{2}\delta_{4}\\
\nonumber ---\\
\nonumber C_3 \delta_{3} C^T_3 && = -\delta_{5}\\
\nonumber C_3 \delta_{5} C^T_3 &&= -\delta_{6} \\
 C_3 \delta_{6} C^T_3 &&= \delta_{3}
\end{eqnarray}
For $\delta_{1}$ pairing, $Tr[m_{[110],\Gamma_8}] = 1$, $Tr[S_4] = 1$ and $Tr[C_3] = 1$. Thus, it belongs to $(A_1)\Gamma_1$ IrRep. For ${\delta_{2,4}}$, $Tr[m_{[110],\Gamma_8}] = 0$, $Tr[S_4] = 0$ and $Tr[C_3] = -1$. Thus, they belong to $E(\Gamma_3)$ Irrep. For $\delta_{3,5,6}$, $Tr[m_{[110],\Gamma_8}] = 1$, $Tr[S_4] = -1$ and $Tr[C_3] = 0$. Thus, they belong to $T_2(\Gamma_4)$ IrRep. The on-site Cooper pairings, written in a more detailed way, are listed in Table \ref{pairf}.

Another possible gap function is constructed by the septet Cooper pairing state\cite{brydon2016}. The gap function is of p-wave state. Thus, we need symmetric 4 by 4 matrix under transpose operation. Since $\{k_x, k_y, k_z\}$ belong to $T_2$ irreducible representation, we can define the following matrices that belong to $T_2$ IrReps
$\Lambda_1=\frac{i}{4}\left(\begin{array}{cccc}
          3&0&0&\sqrt{3}\\
          0&3&\sqrt{3}&0\\
          0&\sqrt{3}&-3&0\\
          \sqrt{3}&0&0&-3
    \end{array}
	\right)$, $\Lambda_2=\frac{1}{4}\left(\begin{array}{cccc}
          3&0&0&-\sqrt{3}\\
          0&-3&\sqrt{3}&0\\
          0&\sqrt{3}&3&0\\
          -\sqrt{3}&0&0&-3
    \end{array}
	\right)$  and $\Lambda_3=\frac{1}{2}\left(
	\begin{array}{cccc}
          0&0&\sqrt{3}&0\\
          0&0&0&\sqrt{3}\\
          \sqrt{3}&0&0&0\\
          0&\sqrt{3}&0&0
    \end{array}
	\right)$.
We find that
\begin{eqnarray}
\nonumber m_{[110],\Gamma_8} \Lambda_1 m^T_{[110],\Gamma_8} &&= -\Lambda_2 \\
\nonumber m_{[110],\Gamma_8} \Lambda_2 m^T_{[110],\Gamma_8} &&= -\Lambda_1 \\
\nonumber m_{[110],\Gamma_8} \Lambda_3 m^T_{[110],\Gamma_8} &&= \Lambda_3 \\
\nonumber ---\\
\nonumber S_4 \Lambda_1 S^T_4 &&= \Lambda_2 \\
\nonumber S_4 \Lambda_2 S^T_4 &&= -\Lambda_1 \\
\nonumber S_4 \Lambda_3 S^T_4 &&=- \Lambda_3 \\
\nonumber ---\\
\nonumber C_3 \Lambda_1 C^T_3 && = \Lambda_2\\
\nonumber C_3 \Lambda_2 C^T_3 &&= \Lambda_3 \\
C_3 \Lambda_3 C^T_3 &&= \Lambda_1
\end{eqnarray}
For $\Lambda_{1,2,3}$, $Tr[m_{[110],\Gamma_8}] = 1$, $Tr[S_4] = -1$ and $Tr[C_3] = 0$. Thus, they belong to $T_2(\Gamma_4)$ IrRep. As a result, we construct the gap function as $\delta_p = k_y\Lambda_1 + k_x\Lambda_2 + k_z \Lambda_3$. One can easily check that $\delta_p$ indeed falls into $A_1(\Gamma_1)$ IrRep. The gap function can be rewritten in the explicit format
\begin{eqnarray}
\delta_p  = \frac{\Delta_p}{4}\left(\begin{array}{cccc}
           3k_+&0&2\sqrt{3}k_z&-\sqrt{3}k_-\\
          0&-3k_-&\sqrt{3}k_+&2\sqrt{3}k_z\\
          2\sqrt{3}k_z&\sqrt{3}k_+&3k_-&0\\
          -\sqrt{3}k_-&2\sqrt{3}k_z&0&-3k_+
    \end{array}
	\right).
\end{eqnarray}

\subsection{Cooper pairings between $\Gamma_6$ and $\Gamma_8$ bands}
We define eight possible linearly independent 2 by 4 matrices as following $\delta_7 =\frac{i}{2}\left(
	\begin{array}{cccc}
          1&0&0&-\sqrt{3}\\
          0&1&-\sqrt{3}&0
    \end{array}
	\right)$, $\delta_8 =\frac{1}{2}\left(
	\begin{array}{cccc}
          1&0&0&\sqrt{3}\\
          0&-1&-\sqrt{3}&0
    \end{array}
	\right)$, $\delta_9 =\frac{i}{2}\left(
	\begin{array}{cccc}
          -\sqrt{3}&0&0&-1\\
          0&-\sqrt{3}&-1&0
    \end{array}
	\right)$, $\delta_{10} =\frac{1}{2}\left(
	\begin{array}{cccc}
          \sqrt{3}&0&0&-1\\
          0&-\sqrt{3}&1&0
    \end{array}
	\right)$, $\delta_{11} =\left(
	\begin{array}{cccc}
          0&1&0&0\\
          1&0&0&0
    \end{array}
	\right)$, $\delta_{12} =\left(
	\begin{array}{cccc}
          0&-i&0&0\\
          i&0&0&0
    \end{array}
	\right)$, $\delta_{13} =\left(
	\begin{array}{cccc}
          0&0&1&0\\
          0&0&0&1
    \end{array}
	\right)$ and $\delta_{14} =\left(
	\begin{array}{cccc}
          0&0&i&0\\
          0&0&0&-i
    \end{array}
	\right)$.

For mirror reflection with respect to (110) plane, we have
\begin{eqnarray}
\nonumber m_{[110],\Gamma_6} \delta_{11} m_{[110],\Gamma_8}^T &&= -\delta_{11} \\
\nonumber m_{[110],\Gamma_6} \delta_{13} m_{[110],\Gamma_8}^T &&= \delta_{13} \\
\nonumber ---\\
\nonumber m_{[110],\Gamma_6} \delta_{7} m_{[110],\Gamma_8}^T &&= -\delta_{8} \\
\nonumber m_{[110],\Gamma_6} \delta_{8} m_{[110],\Gamma_8}^T &&= -\delta_{7} \\
\nonumber m_{[110],\Gamma_6} \delta_{12} m_{[110],\Gamma_8}^T &&= \delta_{12} \\
\nonumber ---\\
\nonumber m_{[110],\Gamma_6} \delta_{9} m_{[110],\Gamma_8}^T &&= \delta_{10} \\
\nonumber m_{[110],\Gamma_6} \delta_{10} m_{[110],\Gamma_8}^T &&= \delta_{9} \\
m_{[110],\Gamma_6} \delta_{14} m_{[110],\Gamma_8}^T &&= -\delta_{14}
\end{eqnarray}

For $S_4$ operation, we have
\begin{eqnarray}
\nonumber S_{4,\Gamma_6} \delta_{11} S_{4,\Gamma_8}^T &&= -\delta_{11} \\
\nonumber S_{4,\Gamma_6} \delta_{13} S_{4,\Gamma_8}^T &&= \delta_{13} \\
\nonumber ---\\
\nonumber S_{4,\Gamma_6} \delta_{7} S_{4,\Gamma_8}^T &&= -\delta_{8} \\
\nonumber S_{4,\Gamma_6} \delta_{8} S_{4,\Gamma_8}^T &&= \delta_{7} \\
\nonumber S_{4,\Gamma_6} \delta_{12} S_{4,\Gamma_8}^T &&= -\delta_{12} \\
\nonumber ---\\
\nonumber S_{4,\Gamma_6} \delta_{9} S_{4,\Gamma_8}^T &&= \delta_{10} \\
\nonumber S_{4,\Gamma_6} \delta_{10} S_{4,\Gamma_8}^T &&= -\delta_{9} \\
S_{4,\Gamma_6} \delta_{14} S_{4,\Gamma_8}^T &&= \delta_{14}
\end{eqnarray}

For $C_3$ operation, we have
\begin{eqnarray}
\nonumber C_{3,\Gamma_6} \delta_{11} C_{3,\Gamma_8}^T &&= -\frac{1}{2}\delta_{11}-\frac{\sqrt{3}}{2}\delta_{13} \\
\nonumber C_{3,\Gamma_6} \delta_{13} C_{3,\Gamma_8}^T &&= \frac{\sqrt{3}}{2}\delta_{11}-\frac{1}{2}\delta_{13} \\
\nonumber ---\\
\nonumber C_{3,\Gamma_6} \delta_{7} C_{3,\Gamma_8}^T &&= \delta_{12} \\
\nonumber C_{3,\Gamma_6} \delta_{8} C_{3,\Gamma_8}^T &&= \delta_{7} \\
\nonumber C_{3,\Gamma_6} \delta_{12} C_{3,\Gamma_8}^T &&= \delta_{8} \\
\nonumber ---\\
\nonumber C_{3,\Gamma_6} \delta_{9} C_{3,\Gamma_8}^T &&= \delta_{14} \\
\nonumber C_{3,\Gamma_6} \delta_{10} C_{3,\Gamma_8}^T &&= \delta_{9} \\
C_{3,\Gamma_6} \delta_{14} C_{3,\Gamma_8}^T &&= \delta_{10}
\end{eqnarray}

Thus, for $\delta_{11}, \delta_{13}$, we have $Tr[m_{[110],\Gamma_8}] = 0$, $Tr[S_{4}] = 0$ and $Tr[C_{3}] = -1$. They fall into 2D $E(\Gamma_3)$ IrRep. For $\delta_{7}, \delta_{8}, \delta_{12}$, we have $Tr[m_{[110],\Gamma_8}] = 1$, $Tr[S_{4}] = -1$ and $Tr[C_{3}] = 0$. They fall into 3D $T_2(\Gamma_4)$ IrRep. For $\delta_{9}, \delta_{10}, \delta_{14}$, we have $Tr[m_{[110],\Gamma_8}] = -1$, $Tr[S_{4}] = 1$ and $Tr[C_{3}] = 0$. They fall into 3D $T_1(\Gamma_5)$ IrRep.

\section{Gap functions in the original basis}
\label{gap_original_basis}
The s-wave gap functions under the \textit{original} basis function $\Psi_{BdG} = (\Psi(\bold{k}),\Psi^{\dag T}(-\bold{k}))^T$ with $\Psi = (\vert \Gamma_6,1/2 \rangle,\vert \Gamma_6,-1/2 \rangle,\vert \Gamma_8,1/2 \rangle,\vert \Gamma_8,-1/2 \rangle,\vert \Gamma_8,3/2 \rangle,\vert \Gamma_8,-3/2 \rangle)^T$ can be written as
\begin{eqnarray}
&&\Delta_{orgn} = \left( 	
\begin{array}{ccc}
a_1&b_1&c_1\\
-b^T_1&a_2&b_2\\
-c^T_1&-b^T_2&a_3
\end{array}
\right)
\end{eqnarray}
where particle-hole(PH) symmetry invairant condition $\Delta_{orgn}(\bold{k}) = - \Delta_{orgn}^T(-\bold{k})$ is used, $a_i$, $b_i$ and $c_1$ are $2\times 2$ matrices and $a_i^T=-a_i$. For s-wave gap functions invariant under time-reversal(TR) symmetry[$\tilde{\mathcal{T}} = \left ( \begin{array}{cc} \mathcal{T} & 0 \\ 0&\mathcal{T}^{\dag T} \end{array} \right)$], one arrives at
\begin{eqnarray}
\nonumber \mathcal{T} \Delta_{orgn} \mathcal{T}^{T} &&= \Delta_{orgn}.
\end{eqnarray}
Moreover, one can divide the s-wave gap functions into two classes according to their behaviors under the mirror symmetry
\begin{equation}
\mathcal{M}_{[110]} \Delta_{orgn} \mathcal{M}_{[110]}^T = \eta \Delta_{orgn},
\end{equation}
where $\eta=\pm 1$ means the gap function is even/odd under the mirror operation.

\subsection{Case: Even under mirror operation}
In this case, the gap function on basis function $\Psi_{BdG}$, based on the previous discussion, is written as

\begin{eqnarray}
\Delta_{orgn}^{(+)} = \left( 	
\begin{array}{ccc}
i\xi_1 \sigma_y & \xi_4\sigma_y +i/\sqrt{2}\xi_5(\sigma_0 + i\sigma_z)& \xi_8 \sigma_0 +i\xi_9/\sqrt{2} (\sigma_x +\sigma_y)\\
\xi_4\sigma_y -i/\sqrt{2}\xi_5(\sigma_0 + i\sigma_z)&i\xi_2 \sigma_y & i\xi_6 \sigma_0 +\xi_7/\sqrt{2} (\sigma_x +\sigma_y)\\
-\xi_8 \sigma_0 +i\xi_9/\sqrt{2} (-\sigma_x +\sigma_y)& -i\xi_6 \sigma_0 +\xi_7/\sqrt{2} (-\sigma_x +\sigma_y)&i\xi_3 \sigma_y
\end{array}
\right)\ ,
\label{pair_m+}
\end{eqnarray}
where $\xi_{1,...,9}$ are real.

Now we come back to the basis function  $\Psi_{BdG} = (c_{+i}(\bold{k}), c_{-i}(\bold{k}), c^{\dag T}_{+i}(-\bold{k}),c^{\dag T}_{-i}(-\bold{k}))^T$ with $c_{\pm i}$ defined in the main text. Only mirror eigenvalue $+i$ subspace is considered and the $-i$ subspace is related to the mirror parity $+i$ subspace by TR symmetry.

The gap functions on the basis function in mirror parity $+i$ subspace, $\Psi^{(+)} = (c_{+i}(\bold{k}), c^\dag_{-i}(-\bold{k}))^T$, is expressed as
\begin{eqnarray}
&&\Delta^{(+)}_{+i} = \left( 	
\begin{array}{ccc}
\xi_1&-i\xi_4+\xi_5&-\xi_8-i\xi_9\\
i\xi_4+\xi_5&-\xi_2&i\xi_6+\xi_7\\
-\xi_8+i\xi_9&-i\xi_6+\xi_7&\xi_3
\end{array}
\right)
\label{gap-m+}
\end{eqnarray}
The resulting Hamiltonian in the mirror parity $+i$ subspace, thus, is expressed as
\begin{eqnarray}
&&H^{(+)}_{+i} = \left( 	
\begin{array}{cc}
e_{+i}(\bold{k})&\Delta^{(+)}_{+i}\\
h.c.&-e^*_{-i}(-\bold{k})
\end{array}
\right)
\end{eqnarray}
where $e_{+i}$ and $e_{-i}$ are $3\times 3$ matrices, defined in Eq. \ref{eq:+i} and \ref{eq:-i}. This Hamiltonian only respects the chiral symmetry $\Pi^{(+)}_{+i}$. Since the TR and PH symmetry under the original basis function $\Psi$ are $\mathcal{T}_{BdG} = \tau_0 \otimes diag[-i\sigma_yK,i\sigma_yK,-i\sigma_yK]$ and $C_{BdG} = \tau_x \otimes \mathcal{I}_{6 \times 6}K$, the chiral symmetry is $\Pi^{(+)}_{BdG} = \tau_x \otimes  diag[-i\sigma_y,i\sigma_y,-i\sigma_y]$. Furthermore, one can obtain the chiral symmetry in the mirror parity $+i$ subspace $\Pi^{(+)}_{+i} = -i\tau_y \otimes\mathcal{I}_{3 \times 3}$. One can do a further transformation such that $\Pi^{(+)'}_{+i} = \tau_z \otimes \mathcal{I}_{3 \times 3}$, where $i$ is ignored before $\tau_z$ without changing the result.

\subsection{Case: Odd under mirror operation}
In this case, the gap function on the original basis $\Psi_{BdG}$ is written as

\begin{eqnarray}
&&\Delta_{orgn}^{(-)} = \left( 	
\begin{array}{ccc}
0 & \eta_1\sigma_x +i\eta_2/\sqrt{2} (\sigma_0-i\sigma_z)& i\eta_5 \sigma_z +i\eta_6/\sqrt{2} (\sigma_x-\sigma_y)\\
-\eta_1\sigma_x -i\eta_2/\sqrt{2} (\sigma_0-i\sigma_z)&0 &\eta_3\sigma_z + \eta_4/\sqrt{2}(\sigma_x - \sigma_y) \\
-i\eta_5 \sigma_z +i\eta_6/\sqrt{2} (-\sigma_x-\sigma_y)&-\eta_3\sigma_z + \eta_4/\sqrt{2}(-\sigma_x - \sigma_y)&0
\end{array}
\right),
\label{pair_m-}
\end{eqnarray}
where $\eta_{1,...,6}$ are real.

The gap function can be further projected into the mirror parity $+i$ subspace, with basis function $\Psi^{(+)} = (c_{+i}(\bold{k}), c^\dag_{+i}(-\bold{k}))^T$,
\begin{eqnarray}
&&\Delta^{(-)}_{+i} = \left( 	
\begin{array}{ccc}
0&\eta_1+i\eta_2&i\eta_5+\eta_6\\
-(\eta_1+i\eta_2)&0&-\eta_3+i\eta_4\\
-(i\eta_5+\eta_6)&\eta_3-i\eta_4&0
\end{array}
\label{chpater4-gap-m-}
\right).
\end{eqnarray}

\section{Correspondence between gap functions $\delta_i$ and $\xi_i\&\eta_i$}
\label{correspondence_delta_xi}
The relationship between these two sets of notations of gap functions are listed in Table \ref{correspondence}.

\begin{center}
\begin{table}[htb]
  \centering
  \begin{minipage}[t]{1.\linewidth}
  \centering
	  \caption{Correspondence between $\delta_i$ and $\xi_i\&\eta_i$.
	  }
\label{correspondence}
\hspace{-1cm}
\begin{tabular}
[c]{c|c|c}\hline\hline
$\delta_i$     &$\xi_i\&\eta_i$ &IrRep\\\hline
$\delta_s$ & $\xi_1$ &$A_1$[$\Gamma_1$]  \\ \arrayrulecolor{green}\hline\arrayrulecolor{black}
$\delta_1$ & $\frac{1}{2}(\xi_2-\xi_3)$&$A_1$[$\Gamma_1$] \\\hline\hline
$\delta_2$ & $\frac{1}{2}(\xi_2+\xi_3)$&\\
$\delta_4$ & $\eta_3$ & E[$\Gamma_3$]\\\arrayrulecolor{green}\hline\arrayrulecolor{black}
$\delta_{11}$ & $\eta_1$&\\
$\delta_{13}$ & $\xi_8$& E[$\Gamma_3$]\\ \arrayrulecolor{black}\hline\hline
$\delta_3$ & $-\xi_6$ &\\
$\delta_5$ & $\frac{1}{\sqrt{2}}[\xi_7 + \eta_4]$ &\\
$\delta_6$ & $\frac{1}{\sqrt{2}}[\xi_7 - \eta_4]$ & $T_2$[$\Gamma_4$]\\\arrayrulecolor{green}\hline\arrayrulecolor{black}

$\delta_7$ & $\frac{1}{2\sqrt{2}}[\xi_5+\eta_2]-\frac{\sqrt{3}}{2\sqrt{2}}[\xi_9+\eta_6]$&\\
$\delta_8$ & $\frac{1}{2\sqrt{2}}[-\xi_5+\eta_2]+\frac{\sqrt{3}}{2\sqrt{2}}[\xi_9-\eta_6]$&\\
$\delta_{12}$ & $\xi_4$&$T_2$[$\Gamma_4$]\\\arrayrulecolor{black}\hline\hline

$\delta_9$ & $-\frac{\sqrt{3}}{2\sqrt{2}}[\xi_5+\eta_2]-\frac{1}{2\sqrt{2}}[\xi_9+\eta_6]$&\\
$\delta_{10}$ &$\frac{\sqrt{3}}{2\sqrt{2}}[-\xi_5+\eta_2]-\frac{1}{2\sqrt{2}}[\xi_9-\eta_6]$&\\
$\delta_{14}$ & $\eta_5$&$T_1$[$\Gamma_5$]\\
\hline\hline
\end{tabular}
  \end{minipage}
\end{table}
\end{center}

\section{D symmetry class of $\delta_3$ gap function for mirror symmetry with respect to (100) planes}
\label{D_for_delta3}
Here we take xz plane as the mirror invariant plane as a concrete example. Since there is full rotational symmetry for the Kane model with vanishing C term, one may check that the mirror symmetry operator with respect to xz plane $\mathcal{M}_y = diag[i\sigma_y,i\sigma_y,-i\sigma_y]$ commutes with the Kane model on the basis $\xi = (\vert \Gamma_6,1/2 \rangle,\vert \Gamma_6,-1/2 \rangle,\vert \Gamma_8,1/2 \rangle,\vert \Gamma_8,-1/2 \rangle,\vert \Gamma_8,3/2 \rangle,\vert \Gamma_8,-3/2 \rangle)^T$. Now we consider the superconducting system with $\delta_3$ gap function, which is written as
\begin{eqnarray}
&&\delta_3 = \left( 	
\begin{array}{ccc}
O&O&O\\
O&O&-i\sigma_0\\
O&i\sigma_0&O
\end{array}
\right)
\end{eqnarray}
where $O$ is a 2$\times$2 matrix with all elements 0. One may check that $\mathcal{M}_y\delta_3\mathcal{M}_y^T = -\delta_3$. In order to obtain the mirror symmetry operator $\tilde{\mathcal{M}}_{y}$ in the Nambu space with the basis $\Phi = (\xi, \xi^{\dag T})$, \textit{i.e.}, $\tilde{\mathcal{M}}_{y} H_{BdG}(k_x,k_z) \tilde{\mathcal{M}}_{y}^{-1} = H_{BdG}(k_x,k_z)$ with $H_{BdG} = \left (\begin{array}{cc}
		H_0(k_x,k_z)&\delta\\
		h.c. &-H_0^*(-k_x,-k_z)
\end{array}\right)$ on the mirror invariant plane (010), we may construct a mirror symmetry operator $\tilde{\mathcal{M}}_{y} = \left (\begin{array}{cc}
		\mathcal{M}_y &\\
		&-\mathcal{M}'^*_y
\end{array}\right)$ for the corresponding BdG Hamiltonian. Obviously, $\{C,\tilde{\mathcal{M}}_y\} = 0$ with $C = \tau_x \otimes \mathcal{I}_{6 \times 6} K$ as the PH symmetry operator, where $\tau_x$ acts on the Nambu space.

Thus, for an eigen wavefunction $\Psi(\bold{k})$ in a mirror subspace with mirror parity $\lambda = \pm i$ on the mirror invariant plane, its PH partner satisfies $\tilde{\mathcal{M}}_y(C \Psi(\bold{k}))  = -C(\tilde{\mathcal{M}}_y \Psi(\bold{k})) = - C(\lambda\Psi(\bold{k})) = \lambda (C\Psi(\bold{k}))$, which indicates that PH symmetry survives in each mirror parity subspace. Namely, the symmetry class in each mirror parity subspace is D.

\end{widetext}

\end{appendix}

\end{document}